\documentclass[letterpaper]{article} 
\usepackage{aaai25}  
\usepackage{times}  
\usepackage{helvet}  
\usepackage{courier}  
\usepackage[hyphens]{url}  
\usepackage{graphicx} 
\usepackage{natbib}  
\usepackage{caption} 
\usepackage{algorithm}
\usepackage{algorithmic}

\usepackage{newfloat}
\usepackage{listings}
\DeclareCaptionStyle{ruled}{labelfont=normalfont,labelsep=colon,strut=off} 
\floatstyle{ruled}
\newfloat{listing}{tb}{lst}{}
\floatname{listing}{Listing}
\usepackage{tabularx}
\usepackage{graphicx}
\usepackage{caption}
\usepackage{subcaption}
\usepackage{color}
\usepackage{xcolor}
\usepackage{multirow}
\usepackage{tcolorbox}
\usepackage{amsmath}
\usepackage{textcomp}

\usepackage{soul}
\sethlcolor{white}

\usepackage{xcolor}
\newcommand{\answerYes}[1]{\textcolor{blue}{#1}} 
 
\newcommand{\answerNA}[1]{\textcolor{gray}{#1}}

\newif\ifstatus
\statustrue

\usepackage{bibentry}

\begin{document}

\title{Cross-Partisan Interactions on Twitter}
\author {
    Yusuf M\"{u}cahit \c{C}etinkaya\textsuperscript{\rm 1,2}*,
    Vahid Ghafouri\textsuperscript{\rm 3,4,5}*,
    Guillermo Suarez-Tangil\textsuperscript{\rm 4},
    Jose Such\textsuperscript{\rm 6,7},\\
    Tuğrulcan Elmas\textsuperscript{\rm 1}
}
\affiliations {
    \textsuperscript{\rm 1}University of Edinburgh\\
    \textsuperscript{\rm 2}Middle East Technical University\\
    \textsuperscript{\rm 3}Oxford Internet Institute\\
    \textsuperscript{\rm 4}IMDEA Networks Institute\\
    \textsuperscript{\rm 5}Universidad Carlos III de Madrid\\
    \textsuperscript{\rm 6}King’s College London\\
    \textsuperscript{\rm 7}Universitat Politecnica de Valencia\\
    yusufc@ceng.metu.edu.tr, vahid.ghafouri@oii.ox.ac.uk, guillermo.suarez-tangil@imdea.org, jose.such@kcl.ac.uk, telmas@ed.ac.uk
}

\maketitle

\def\thefootnote{*}\footnotetext{These authors contributed equally.}

\begin{abstract}
Many social media studies argue that social media creates echo chambers where some users only interact with peers of the same political orientation. However, recent studies suggest that a substantial amount of Cross-Partisan Interactions (CPIs) do exist --- even within echo chambers, but they may be toxic. There is no consensus about how such interactions occur and when they lead to healthy or toxic dialogue. In this paper, we study a comprehensive Twitter dataset that consists of 3 million tweets from 2020 related to the U.S. context to understand the dynamics behind CPIs. We investigate factors that are more associated with such interactions, including how users engage in CPIs, which topics are more contentious, and what are the stances associated with healthy interactions. We find that CPIs are significantly influenced by the nature of the topics being discussed, with politically charged events acting as strong catalysts. The political discourse and pre-established political views sway how users participate in CPIs, but the direction in which users go is nuanced. While Democrats engage in cross-partisan interactions slightly more frequently, these interactions often involve more negative and nonconstructive stances compared to their intra-party interactions. In contrast, Republicans tend to maintain a more consistent tone across interactions. Although users are more likely to engage in CPIs with popular accounts in general, this is less common among Republicans who often engage in CPIs with accounts with a low number of followers for personal matters. Our study has implications beyond Twitter as identifying topics with low toxicity and high CPI can help highlight potential opportunities for reducing polarization while topics with high toxicity and low CPI may action targeted interventions when moderating harm. 
\end{abstract}

\section{Introduction}

Social media is often criticized for creating \emph{echo chambers}, where individuals are primarily exposed to and interact with content that aligns with their beliefs~\cite{barbera2015birds}.
Such spaces reinforce existing views, limit exposure to diverse perspectives, and contribute to polarization. 
However, recent studies argue that there is a significant amount of \emph{Cross-Partisan Interactions}, CPIs, engagements among users with different political views, challenging the notion that social media only deepens divides~\cite{rajadesingan2021walking}. 

CPIs hold significance on social media as they serve as potential bridges between divided ideological groups, facilitating exposure to diverse perspectives and fostering dialogue among disparate communities. 
Meanwhile, there is a debate on whether such interactions lead to healthy dialogue and mitigate polarization, or are toxic and exacerbate it. 
Such contrasting opinions suggest that the social media dynamics that affect the presence and quality of CPIs are complex. 
Understanding these dynamics is crucial to understanding CPIs and subsequently, how information and differing viewpoints propagate through networks, potentially influencing societal polarization and discourse quality. 
This broadened exposure can lead to a more informed populace and potentially temper the polarization that is rampant in online spaces. 
Furthermore, understanding CPIs is crucial for platform designers and policymakers who aim to create healthier online environments. 

Our main contribution is investigating the dynamics of CPIs by identifying the interplay between actors, topics, and stances that are more associated with CPIs. To do so, we conduct the first large-scale topic-independent quantitative analysis of Cross-Partisan Interactions to the best of our knowledge.
We also study in which cases the CPIs are \emph{healthy}, i.e., non-toxic. 
For this, we formulate the following Research Questions (RQs): 

\vspace{.1cm}
\noindent\textbf{RQ1:} What are the characteristics of {\bf users} who typically engage with (un)healthy cross-partisan interactions? 

\vspace{.1cm}
\noindent\textbf{RQ2:} Which {\bf topics} are more prevalent in healthy or unhealthy cross-partisan interactions?

\vspace{.1cm}
\noindent\textbf{RQ3:} What type of {\bf content}, in terms of sentiment and stance, appears in healthy cross-partisan interactions?

\noindent We collect a comprehensive reply dataset and devise a methodology to determine the political alignment of users, and analyze the stance of toxicity of the content (\S\ref{sec:method}) to answer these questions (\S\ref{sec:results}-\ref{sec:content-analysis}). While doing so, we find that popular users are more likely to foster CPIs as they attract replies from users with opposite views. In contrast, the repliers in CPIs tend to have lower number of followers, followings and tweet counts. Certain topics such as politics, economics, and golf has higher rates of CPI in contrast to public health, religion, and daily life. CPIs are correlated with heightened toxicity, which suggests that while these interactions can bridge ideological gaps, they also risk escalating tensions. Furthermore, the study confirms that more polarized users tend to use offensive language. We discuss the findings in (\S\ref{sec:discussion}).

\section{Related Work}
\label{sec:related-work}

Cross-partisan interactions (CPIs) can potentially create bridges in polarized political environments and disrupt echo chambers. In juxtaposition, they may also exacerbate existing divides and increase toxicity online. We now provide a brief survey on these research themes.

\noindent\textbf{Echo Chambers:} Echo chambers are environments where people are exposed only to information and opinions that reinforce their existing beliefs, leading to a lack of exposure to diverse perspectives and the potential for increased polarization. Past studies argue for the existence of echo chambers online by employing user data ~\cite{garimella2018political, elmas2021can, torres2022manufacture}. However, some studies suggest that their prevalence and impact may be overstated ~\cite{dubois2018echo,guess2018selective}, 
A substantial portion of the audience accesses news from outlets with varying ideological perspectives from both online and offline platforms~\cite{gentzkow2011ideological}. 
Such cross-partisan media consumption indicates that individuals are not wholly isolated within echo chambers but rather encounter diverse viewpoints. 

\noindent\textbf{Cross-Partisan Interactions:} The impact of cross-partisan interactions on polarization and dialogue health is a subject of ongoing debate. Past studies have not reached a consensus on whether such interactions bridge divides or exacerbate tensions, e.g.: \citeauthor{bail2018exposure} (\citeyear{bail2018exposure}) and \citeauthor{rajadesingan2021walking} (\citeyear{rajadesingan2021walking}) suggest cross-partisan interaction increases polarization, while works like \citeauthor{rajadesingan2021walking} (\citeyear{rajadesingan2021walking}) and \citeauthor{ccetinkaya2024towards} (\citeyear{ccetinkaya2024towards}) highlight the potential for CPIs to foster understanding and reduce polarization.  

Others show that the impact of cross-partisan interactions on polarization is more complex. ~\citeauthor{balietti2021reducing}~(\citeyear{balietti2021reducing}) find that exposure to cross-partisan content can decrease polarization if the exposed feel close to the author. ~\citeauthor{marchal2022nice}~(\citeyear{marchal2022nice}) report that negative cross-partisan interactions increase political polarization, but positive ones mitigate it on Reddit.~\citeauthor{zade2024reply}~(\citeyear{zade2024reply}) reveal that the reply feature is often used to re-frame and shift the attention to different dimensions of the debate that were not salient in the original tweet, which may or may not exacerbate tensions.

These studies suggest that the dynamics of cross-partisan interactions are complex, and their impact varies on a case-by-case basis. Our study adopts a comprehensive approach to understanding which actors and topics are more associated with cross-partisan interactions and on which occasions such cross-partisan interactions are more toxic. To the best of our knowledge, such a case-by-case understanding of the likeliness and impact of CPIs is not clearly established.

Perhaps the most similar work to ours, by ~\citeauthor{an2024curated} (\citeyear{an2024curated}), that uses COVID-related Twitter data suggests that Republicans were more likely to initiate cross-party interactions, but such interactions were driven by a small set of users. We extend this work by employing a general dataset that stems from the 1\% sample of Twitter instead of limiting it to Covid so that we can analyze which actors and topics are more likely to create cross-partisan interactions extensively. We also propose an unbounded analysis using LLMs to understand the prevalent stances in the data.

\citeauthor{shugars2019keep} (\citeyear{shugars2019keep}) employs a dataset of tweets related to Trump and proposes a model that predicts user participation in prolonged conversations. Our problem differs from theirs as they predict if a given user will reply to a conversation thread while we try to predict the factors that are more associated with cross-partisan interactions.

Finally, ~\citeauthor{santoro2022promise} (\citeyear{santoro2022promise}) find that the effectiveness of cross-partisan conversations depends on the topic through a randomized controlled trial. They employed two topics of discussion: a non-political topic discussing the perfect day, and the other discussing their favorite political party. The former mitigates the polarization, while the other has no effect. We complement this work by testing their hypothesis in practice and employing a unique, comprehensive social media analysis that is not limited to only two topics.

\noindent\textbf{Text-mining on Online Interactions:} To analyze user interactions at scale, social media studies employ text-mining methods such as detecting the sentiment, stance, toxicity, or discourse act of the content, employing classification using a predefined set of classes, or assigning a score.

For instance, to analyze the online debate on the Italian 2016 referendum \citeauthor{Lai2019italystance}~(\citeyear{Lai2019italystance}) use a semi-supervised approach by manually annotating a corpus of Italian tweet-replies as in ``\textit{FAVOR}'', ``\textit{AGAINST}'', or neutral (``\textit{NONE}'') to the root tweet, and training a SVM classifier on the labels using \textit{bag-of-words}, \textit{bag-of-hashtags}, and \textit{bag-of-mentions} as features. The labels they annotate by such prediction indicate that users use replies more frequently to express divergent opinions than other communication types. \citeauthor{Matalon2021israelsentiment}~(\citeyear{Matalon2021israelsentiment}) employ VADER~\cite{hutto2014vader} to evaluate the sentiment of tweets related to Israel to obtain a sentiment score in the range: [-1, +1]. The results indicate that a root tweet's sentiment influences the likelihood of receiving a quote that inverts its opinion. 

While past studies mainly employ lexicon based~\cite{hutto2014vader} or transformer-based~\cite{karande2021stance} approaches, recent studies suggest that LLMs are capable of text mining tasks such as \textit{stance detection}~\cite{Lan2024LLMstance}, \textit{sentiment analysis}~\cite{lee2024masking}, \textit{hate speech}~\cite{kaya2024arc}, and sociopolitical affiliation of texts~\cite{ghafouri2023kialo}. 

In our work, we employ two text-mining methods: unbounded content annotation and toxicity detection. We employ the former to analyze the stance and sentiment of the interactions qualitatively. We opt for this method to go beyond predefined sentiment and stance classes and understand nuances in online interactions, which is novel. For toxicity detection, we use the transformers-based Perspective API~\cite{lees2022new} which provides a toxicity score for a given text between 0 and 1. Prior research identifies a threshold to classify if a given social media post is toxic or not. For instance, \citeauthor{kumar2021designing} (\citeyear{kumar2021designing}) \hl{survey 17,280 participants to rate toxic content. They then identify a personalized threshold for each participant. They find that the median optimal threshold across participants 0.61.} Similarly, \citeauthor{aleksandric2024users} (\citeyear{aleksandric2024users}) \hl{reports that Perspective API achieves the highest accuracy of 0.95 at a threshold of 0.6 for the ``Toxicity'' attribute on a Twitter dataset they annotated.}

\section{Data \& Methodology}
\label{sec:method}

We define a \emph{cross-partisan interaction}, CPI for short, as a direct interaction between two users of different political orientations. We use Twitter (X) as the platform to study. Twitter features four types of interactions between users: retweets, likes, quotes, and replies. We limited our analysis only to replies, as only replies provide evidence that people intend to engage in a direct reciprocal interaction that can lead to a dialogue~\cite{an2024curated,zade2024reply}. 

To study CPIs, we first collect a dataset of interactions in the form of replies, replied tweets, and root tweets. We then employ political orientation detection to identify partisanship and discover cross-partisan interactions.

We limit our focus to the U.S. context and define partisanship as left-aligned (leaning towards liberals or Democrats) or right-aligned (leaning towards conservatives or Republicans). We limit the data period to 2020 as it captures the general discussions, such as the pandemic (often non-political) and the political discussions related to the 2020 U.S. presidential election. The dataset and related labels used in the analysis are shared on GitHub.\footnote{\url{https://github.com/tweetpie/cross-partisan-interactions}.}

\noindent\textbf{Replies \& Roots:} To have an unbiased sample of replies, we employed the 1\% random sample of Twitter provided by the Internet Archive~\cite{team2020archive}. We recollected the replies and the tweets they replied to in March 2023 using the Twitter API. The dataset comprises 3,029,231 reply tweets in English, responding to 2,299,444 unique tweets. However, on Twitter, tweets can be part of a reply chain. To simplify the analysis, we discard the nested replies and limit the analysis where the replied tweet is not a reply of the original tweet (namely, \emph{root}). This brings the dataset to 1,870,794 direct replies (61.8\% of all replies), replying to 1,175,151 root tweets. There are 683,450 unique repliers and 211,584 root authors. \hl{The dataset is further filtered by only keeping the users with non-neutral political orientation, using the methodology that we will describe now.}

\noindent\textbf{Political Orientation:} We employ the methodology of \citeauthor{barbera2015birds}~(\citeyear{barbera2015birds}) to measure users' political orientation. The method uses Bayesian inference on users following data to assign a political orientation score to them, which ranges between -5 and +5. Negative values signify leaning toward Democrats and positive values mean leaning toward Republicans. Our dataset with only direct replies contains 821,791 users. Among these, 57,037 users (7.0\%) are not assigned a score due to the absence of the following data and are excluded from the analysis. We see 493,954 \hl{(60.1\%)} users classified as left-aligned and 232,186 \hl{(28.3\%)} classified as right-aligned due to having an absolute score above 0.1. There are 38,614 users with a score between -0.1 and 0.1, that are considered neutral and discarded from the analysis.

\hl{The direct replies and root tweets from users with unassigned political orientations and from neutral users are discarded from the dataset and are not used in the rest of the analysis. The final dataset comprises 2,561,846 tweets from 726,140 users, including 1,594,271 replies from 596,749 users and 967,575 root tweets from 142,219 users.} 

\noindent\textbf{CPI Data:} We classify an interaction as a CPI if the replier and the root author are assigned a different political orientation. There are 661,661 replies classified as CPI (34\%). Of these, 196,642 are from Republicans replying to Democrats, making up 33.2\% of all Republican tweets, and 432,004 are from Democrats replying to Republicans, accounting for 34.3\% of all Democrat tweets. While cross-partisan interactions originating from Democrats constitute 65\% of all CPIs, this is due to the dataset containing a larger number of users labeled as Democrats. 

\noindent\textbf{Toxic Language Classification:} To determine whether the interactions are healthy or toxic, we employ Google Jigsaw’s Perspective API. It is a public tool based on a transformer model that assigns a score which indicates the toxicity of the language within the text \cite{lees2022new}. The API outputs scores ranging from 0 to 1, indicating the presence of various forms of hate speech, including ``\textit{Toxicity},'' ``\textit{Severe Toxicity},'' ``\textit{Insult},'' ``\textit{Threat},'' ``\textit{Profanity},'' and ``\textit{Identity}''. We streamline the process for our analysis by assigning each tweet the maximum score across these categories. We define a threshold over the score that determines whether a tweet contains offensive language. We set the threshold at 0.61 which is identified by ~\citeauthor{kumar2021designing} (\citeyear{kumar2021designing}) by employing a survey on 17,280 participants and ~\citeauthor{aleksandric2024users} (\citeyear{aleksandric2024users}) who found that Perspective API gets the highest accuracy on this threshold on their independent test set.

\noindent\textbf{Sentiment \& Stance Annotation:}To better understand user interactions, we characterize tweets with annotations describing their sentiment and stance. We use these annotations to perform a comparative content analysis. Due to the large size of our dataset, we resort to automated characterization mechanisms. In particular, we leverage state-of-the-art LLMs. We prompt the root tweets and replies to an LLM and task the model to describe them using three adjectives.
For the replies, the task is to describe the \textit{stance} against the root tweet with three adjectives while we offer both the reply and the root tweet to the LLM. Since root tweets are not usually directed to another tweet, we ask the LLM also to annotate its \textit{sentiment}. This method is an alternative to constraining LLMs by predefined classes and helps us qualitatively analyze the sentiment and stances of the tweets.

Table~\ref{tab:prompt-sample} (in Appendix~\ref{sec:appendix}) provides sample prompts passed to LLM and the completion provided by LLM.  
Later, in Section~\ref{sec:discussion:limitation} we discuss the reason and limitations of our choice of prompt engineering and possible future configurations.

The interactions we characterize in our dataset are of four categories: Democrats replying to Democrats (D\textrightarrow D), Republicans replying to Republicans (R\textrightarrow R), Republicans replying to Democrats (R\textrightarrow D), and Democrats replying to Republicans (D\textrightarrow R).
As, for a fair comparison, we intend to have a balanced amount of annotations from each interaction type, we randomly sample 100,000 (the approximate size of the smallest category of interaction) tweet-reply pairs resulting in a total of 400,000 instances.

We employ ``\textit{Mistral-7B-Instruct-v0.2}''\footnote{\url{https://huggingface.co/mistralai/Mistral-7B-Instruct-v0.2}}.
This open-source model allows for efficient processing of the large dataset while maintaining adequate language-understanding capabilities. We downloaded the model locally from the Hugging-Face. Each query completion took \~5 seconds on an NVIDIA A100 80GB GPU. Due to the light size of the model, we parallelized the process into 10 folds and obtain the answers in \~10 days.

We validate the annotations by manually inspecting a random sample of 100 AI-annotated root tweet \& reply pairs and labeling them as correct or incorrect. \hl{Two annotators independently performed this task} The LLM (\textit{Mistral-7B-Instruct-v0.2}) achieves an accuracy of 97\% and 88\% (as evaluated by annotator 1 and annotator 2, respectively) for determining the sentiments of root tweets, and 88\% and 85\% for identifying the stances of replies toward root tweets. The calculated Kappa inter-annotator agreement scores are 0.26 for the sentiments of root tweets, and 0.44 for the stances of replies toward root tweets, which correspond to fair agreement and moderate agreement respectively~\cite{landis1977kappa}. The former score is lower due to high prevalence of data points annotated as true.
Although not classified as incorrect, we observe that LLMs sometimes annotate the sentiment of the reply instead of its stance, e.g., ``\textit{happy}'' instead of ``\textit{happy for them}'' or ``\textit{supportive}.'' We do not correct these annotations and leave them as a limitation.

\section{RQ1: User Analysis}
\label{sec:results}

This section investigates the characteristics of users who engage in CPIs, addressing RQ1. We first focus on profile attributes to identify the individuals most commonly associated with CPIs. We then adopt a user-specific approach, analyzing each user. We use regression to understand the correlation across features. 

\subsection{Profile \& Tweet Attributes vs. CPI Rates}
Certain profile and tweet attributes may be more likely to foster CPIs. Thus, we investigate the prevalence of certain user-based attributes in the promotion of CPIs. In particular, we look at the following profile and tweet attributes:
(1) number of followers, (2) followings, (2) total tweet count, and (3) tweets' replies, (4) retweets, (5) quote, and (6) like count. We also incorporate the same attributes (1-3) to understand if they have a larger influence than the root user. 

\noindent\textbf{Fundamentals of the Analysis:} We use Logistic Regression to compute the coefficients for these attributes to identify their effect, using 20\% of the dataset as the test set to measure the models' performances. \hl{We train three models: one using the entire dataset, one focusing solely on replies from Democrats, and another focusing on replies from Republicans.}
The dependent variable \( Y \) is whether a tweet is part of a cross-partisan conversation (\( Y = 1 \)) or not (\( Y = 0 \)). The logistic regression model can be expressed as follows: 

\begin{equation}
    \text{logit}(P(Y=1)) = \log\left(\frac{P(Y=1)}{1-P(Y=1)}\right) = \beta_0 + \sum_{i=1}^{n} \beta_i X_i, 
\end{equation}

\noindent where:
\begin{itemize}
    \item \( \beta_0 \) is the intercept of the model.
    \item \( \beta_i \) are the coefficients associated with the \( n \) predictor variables \( X_i \).
    \item \( X_i \) represent the various author-related metrics included in the model.
\end{itemize}

\begin{figure}[ht]
        \centering
        \includegraphics[width=\linewidth]{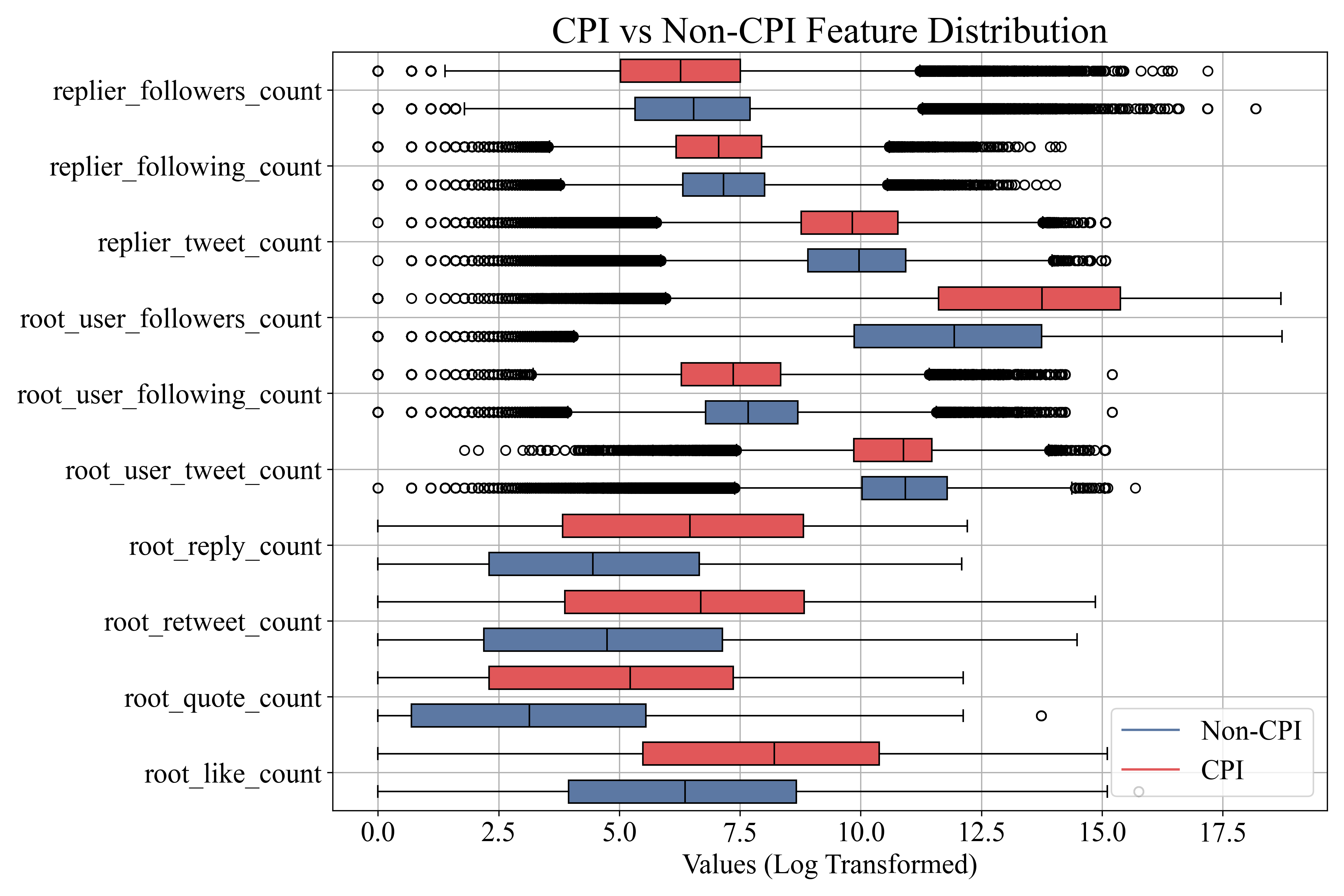}
        \caption{The distribution of the attribute values among the CPIs and non-CPIs.}
        \label{fig:features-boxplot}    
\end{figure}

The logistic regression model \hl{for the entire dataset} is estimated using Maximum Likelihood Estimation (MLE) on a dataset of 1,902,704 observations. Figure~\ref{fig:features-boxplot} displays the distribution of attributes used in the model. The model converges after 6 iterations with a Pseudo R-squared of 0.05413. This indicates that the model explains about 5.4\% of the variance with the given predictors in the outcome variable, which is typical for models in complex social or behavioral datasets where many factors are at play. The model's ROC AUC score is 0.656, which suggests a moderate discriminative ability. \hl{The model that uses Democrats subset achieves a significantly higher Pseudo R-squared value of 0.1598, indicating that the predictors explain about 15.98\% of the variance, compared to only 3.48\% in the Republican model. Similarly, the ROC AUC score for Democrats is 0.6357, reflecting moderate discriminative ability, while the Republican model’s ROC AUC score is only 0.538.} Table~\ref{tab:logit_results} presents the coefficients for each predictor variable \hl{for three models}. All coefficients are statistically significant ($p < 0.0001$).

\begin{table}[ht!]
    \centering
    \footnotesize
\begin{tabular}{lrrr}
\hline
\textbf{Variable} & \textbf{All} & \textbf{Dem.} & \textbf{Rep.} \\
\hline
\texttt{const} & -0.656 & -0.482 & -0.672 \\
\texttt{replier$_{followers\_count}$} & -0.025 & \underline{-0.158} & \underline{0.020} \\
\texttt{replier$_{following\_count}$} & -0.058 & -0.077 & -0.061 \\
\texttt{replier$_{tweet\_count}$} & -0.065 & -0.082 & -0.028 \\
\texttt{root\_user$_{followers\_count}$} & 0.284 & \underline{0.817} & \underline{-0.356} \\
\texttt{root\_user$_{following\_count}$} & -0.020 & \underline{0.032} & \underline{-0.119} \\
\texttt{root\_user$_{tweet\_count}$} & -0.074 & -0.092 & -0.019 \\
\texttt{root$_{reply\_count}$} & 0.525 & \underline{1.978} & \underline{-0.304} \\
\texttt{root$_{retweet\_count}$} & 0.276 & 0.719 & 0.220 \\
\texttt{root$_{quote\_count}$} & -0.199 & -0.378 & -0.164 \\
\texttt{root$_{like\_count}$} & -0.292 & \underline{-1.450} & \underline{0.788} \\
\hline
\end{tabular}
    \caption{Logistic Regression results for CPIs. Coefficients are underlined if the sign is different in two camps.}
    \label{tab:logit_results}
    \footnotesize
\end{table}

\noindent\textbf{Key Predictors:} We observe that accounts fostering CPIs by replying have fewer count of followers, following, and tweets as these have negative coefficients. This means that new or small accounts are more likely to engage in CPIs, possibly to attract attention and grow their accounts. Alternatively, people may be using their fake accounts to engage in CPIs. The differences among the two type of repliers are rather small as in Figure~\ref{fig:features-boxplot}. On the other hand, root tweets from users with high follower count are more associated with CPIs from Democrats but less with CPIs from Republicans. Root users' tweet counts have negative association while their following counts have contrasting but negligible associations with CPIs. We observe that the coefficients for the root's attributes are higher than the repliers' attributes. This suggests that the CPIs depend mostly on root users.

Interestingly, the attributes of root tweets have contrasting effects. The reply count of the root tweet has the biggest coefficient by far for the entire dataset. This suggests a rich-get-richer effect, i.e., popular tweets with many replies also attract many CPIs. This findings is even more pronounced for Democrats, but does not generalize to Republicans, which has a negative coefficient for the same attribute. The tweet's retweet count has a similar association and generalizes to Republicans although the coefficient is low. Quote count is less associated with CPIs for both camps, possibly because users tend to quote tweets without replying to them, thereby reducing CPIs in the form of replies.

Surprisingly, like count has a very high negative coefficient for Democrats and a very high positive coefficient for Republicans. This indicates that Republicans are more inclined to engage in CPIs with highly liked tweets from the opposing camp, whereas Democrats may tend to create discussions on tweets they (and the general public) dislike. Alternatively, the likes may amplify the content in the users' echo chamber, while retweets may go beyond and create cross-partisan interactions. Note that since the data is from 2020, the disparity between these attributes may have changed due to changes in Twitter's algorithms. 

The differences between the camps underline how cross-partisan engagement is influenced not only by root tweet attributes but also by the alignment of the repliers, reflecting distinct behavioral tendencies and engagement strategies.

We would like to clarify that the attributes of root tweets with negative coefficients (e.g., quote and like count for the entire dataset) fostering CPIs are still higher as seen in Figure~\ref{fig:features-boxplot}, since they are correlated with other attributes, e.g., a tweet with many likes and quotes also receive many quotes and replies. However, these attributes contribute negatively to the model, i.e., likes and quotes may be lower for the tweets with comparable retweet and reply counts in CPIs.

\subsection{User-Specific CPI Rates}

While user profile attributes partly explain the type of users that are more associated with CPIs, it does not provide the whole picture. For instance, popular users may differ in their likelihood of initiating CPIs. For a more in-depth analysis, we focus on identifying the key predictors among users. 

\noindent\textbf{Fundamentals of the Analysis:} We employ Lasso regression to model these users. This method is particularly suited to our dataset, which is characterized by high dimensionality and sparsity, i.e., there are many users but a limited number of interactions among them, which is typical of social media \cite{tibshirani1996regression, emmert2019high}. 

We prepare a sparse matrix where rows represent interactions, columns represent users, and each entry is a binary indicator denoting whether the interaction is associated with a specific user. The output variable is binary, with a value of 1 indicating the conversation is CPI and -1 indicating non-CPI. We apply a threshold to eliminate noise from users with only a few interactions, considering only those users who acted as a replier or a root more than 10 times. The filtered dataset has 1,702,019 interactions created by 72,923 users. 24,729 of these users were exclusively repliers, 16,407 users were exclusively roots, and 31,787 users interacted both ways. 

The Lasso regression model is defined as follows:

\begin{equation}
\hat{\beta} = \underset{\beta}{\text{argmin}} \left\{ \frac{1}{N} \sum_{i=1}^{N} (y_i - \mathbf{x}_i^T \beta)^2 + \lambda \sum_{j=1}^{p} |\beta_j| \right\},
\end{equation}

\noindent where \( y_i \) is the response variable, \( \mathbf{x}_i \) is the vector of predictors for the \( i \)-th observation, \( \beta \) is the vector of coefficients, and \( \lambda \) is the regularization parameter that controls the degree of shrinkage applied to the coefficients \cite{friedman2010regularization}. Users with non-zero coefficients are deemed significant contributors to the CPIs. This feature selection process is intrinsic to the Lasso technique and aids in simplifying the model by removing irrelevant predictors.

\begin{figure}[!htb]
        \centering
        \includegraphics[width=0.8\linewidth]{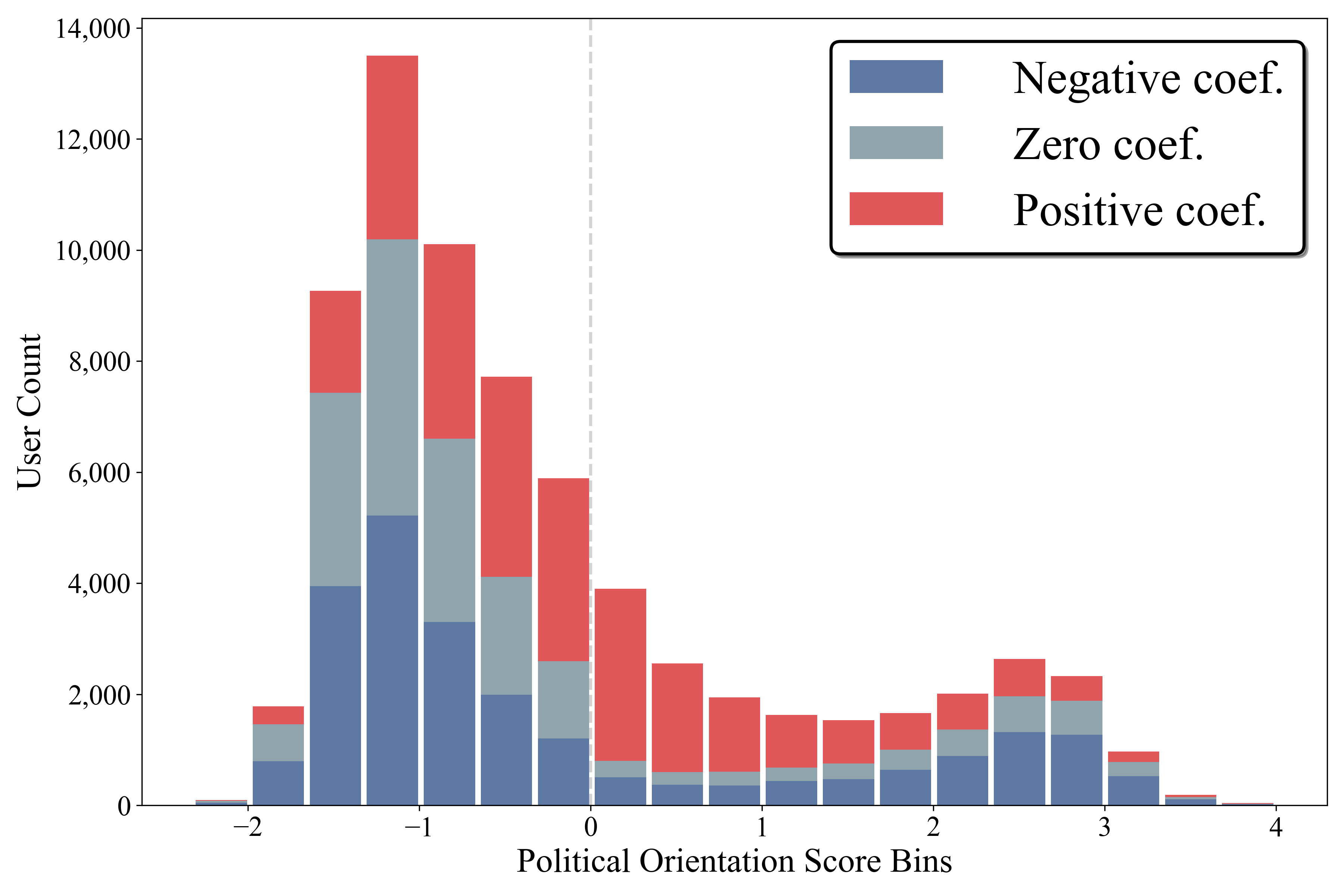}
        \caption{The distribution of political orientation scores for Lasso coefficients of authors on their CPI involvement.}
        \label{fig:coeffs-distro}    
\end{figure}

\noindent\textbf{Key Predictors:} Figure \ref{fig:coeffs-distro} shows the political orientation scores of the authors with their coefficient signatures. There are 26,581 users with negative, 26,649 users with positive, and 19,693 users with zero coefficients.

\begin{figure}[ht]
        \centering
        \includegraphics[width=0.8\linewidth]{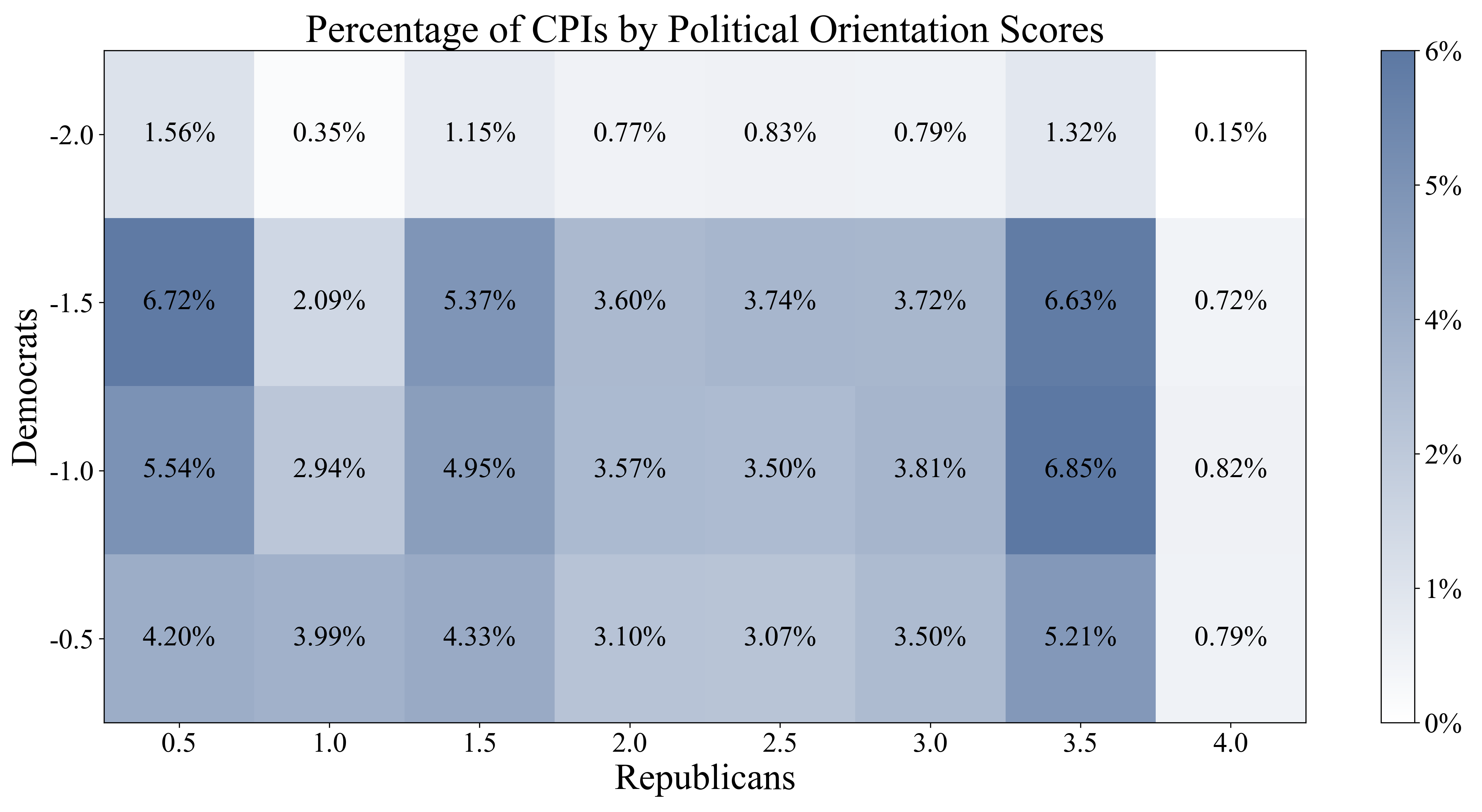}
        \caption{The distribution of CPIs by authors' political orientation scores, grouped into 0.5 interval bins. }
        \label{fig:cpi-replies-heatmap}    
\end{figure}

The coefficients of the users show a Pearson correlation of 0.78 with their CPI versus non-CPI ratios, indicating a strong positive relationship. This suggests that as the coefficient values increase, the likelihood of a user contributing to CPIs also rises proportionally. Our analysis reveals that Democrats tend not to be involved in CPIs, as their bins have more negative coefficient portions compared to Republicans, who are more likely to engage in CPIs. Moreover, the political orientation scores show an inverse relationship with the probability of creating CPIs; users close to the political center have a higher percentage of CPI involvement, which can be interpreted as being in more moderate positions. In contrast, those with more extreme scores on the political spectrum are less likely to be involved in CPIs. We further document this in Figure~\ref{fig:cpi-replies-heatmap}, which shows the distribution of CPIs with respect to political orientation scores.

\begin{figure}[ht]
        \centering
        \includegraphics[width=0.9\linewidth]{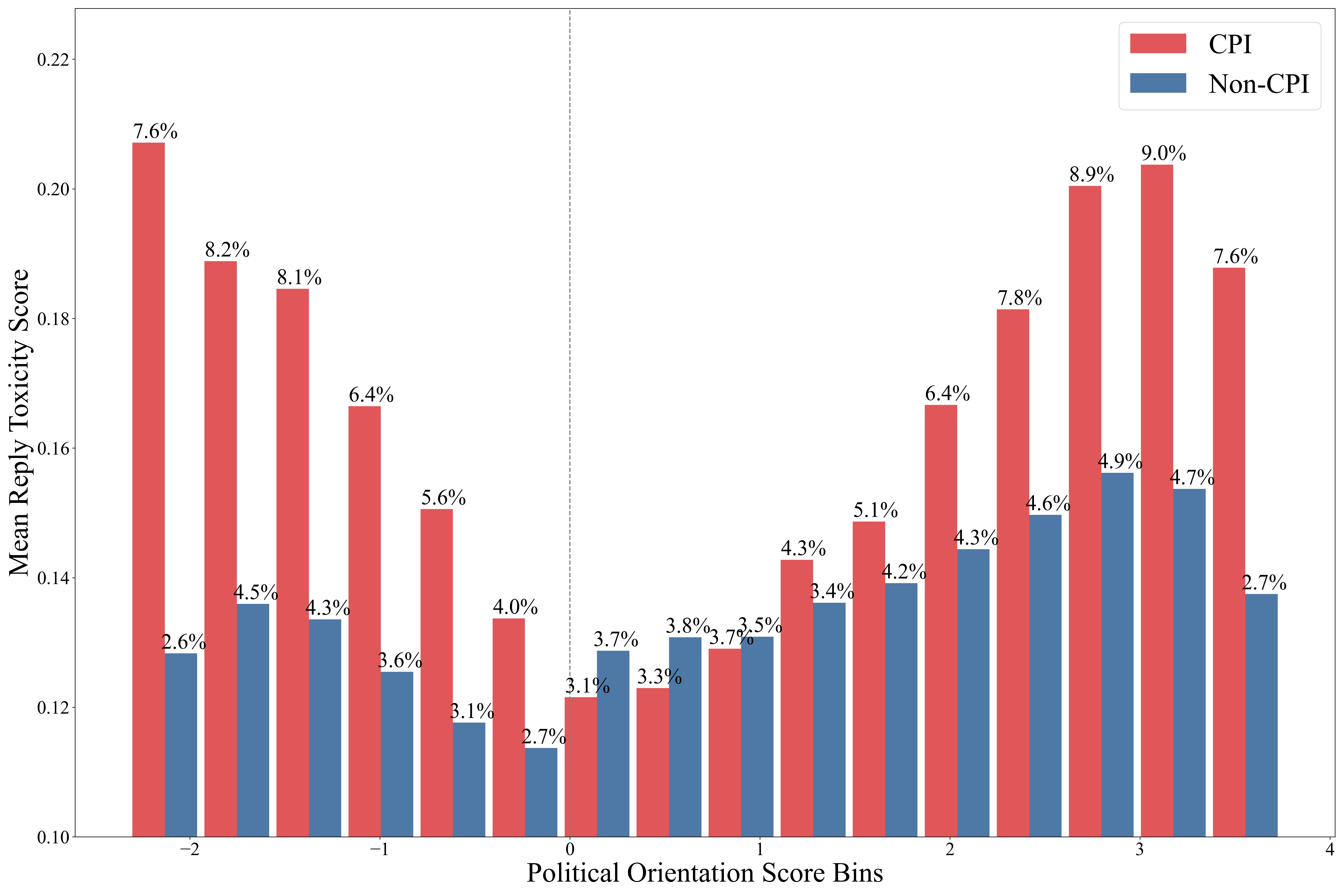}
        \caption{Mean toxicity levels of the repliers' political orientation score bins for both CPI and non-CPIs. The numbers on bars show the percentage of the content that is counted as toxic ($score>0.6$) in each bin.
        }
        \label{fig:reply-toxicity-by-orientation}    
\end{figure}

\noindent\textbf{CPI Toxicity:} We categorize the replies into bins based on the political orientation of the replier. \hl{For each bin, we calculate the average toxicity score of the replies. Additionally, we determine the percentage of toxic tweets in each bin for interpretability, using a toxicity threshold of 0.61. We removed the bins where the replier has a political orientation score greater than 3.5 as there were only 205 replies in total. We repeat this analysis by averaging the reply toxicity per user, so that each data point in a bin represents a user rather than a tweet. Since the results do not differ significantly, we visualize only the first experiment. Furthermore, we tested the sensitivity of our findings by varying the toxicity threshold from 0.4 to 0.9. This adjustment yielded no substantial change in the behavior of the toxicity distribution.} Figure~\ref{fig:reply-toxicity-by-orientation} illustrates the mean reply toxicity scores (bars) and percentage of toxic tweets (bar labels) across different political orientation score bins. The overall trend shows that toxicity levels for CPI replies (red bars) are consistently higher than for non-CPI replies (blue bars) as the political orientation score moves away from the center. As moving toward the center, toxicity scores gradually decrease for both CPI and non-CPI interactions, reaching their lowest around the center, where CPI toxicity drops to 3.1\% and non-CPI toxicity to 2.7\%. Interestingly, toxicity decreases at the edges of the spectrum. Overall, the toxicity scores have a symmetric behavior with respect to the point zero for both parties.

\section{RQ2: Topic Analysis}
\label{sec:rq2}

To answer RQ2, we investigate the characteristics of the topics that are likely to foster CPIs.
We employ BERTopic \cite{grootendorst2022bertopic} pipeline for topic modeling to identify the topics. We first generate dense vector representations of the texts using BERT, a pre-trained transformer model. We then reduce the dimensionality of these embeddings using UMAP to make the clustering more efficient. Finally, we apply HDBSCAN, a clustering algorithm, to group similar embeddings, forming distinct topics. We process ``root tweets'' to generate embeddings for further analysis.

The topic modeling identifies 177 distinct topics within the dataset. We further categorize these topics to ease comprehension and visualization. We utilize the top 10 keywords provided by \texttt{BERTopic}, along with the 10 closest tweets to the centroid of each topic, to create headline summaries and human-readable titles for each topic using ChatGPT-4o. We manually group the topics into 11 overarching \emph{categories} using these summaries to simplify the analysis. \hl{Political topics comprise approximately 35\% of these 177 topics, encompassing discussions on governance, civil rights, election campaigns, and policy debates. A notable subset of these, around 10-15\%, specifically pertains to election-related discourse, reflecting the temporal and sociopolitical context of the data. Non-political topics, by contrast, encapsulate a broad spectrum of interests, including cultural reflections, entertainment, and everyday experiences, such as musical themes, food discussions, and personal milestones.}

\subsection{Topics vs. CPI Rates}

\begin{figure}[!htb]
        \centering
        \begin{subfigure}[b]{\linewidth}
        \centering
        \includegraphics[width=0.8\linewidth]{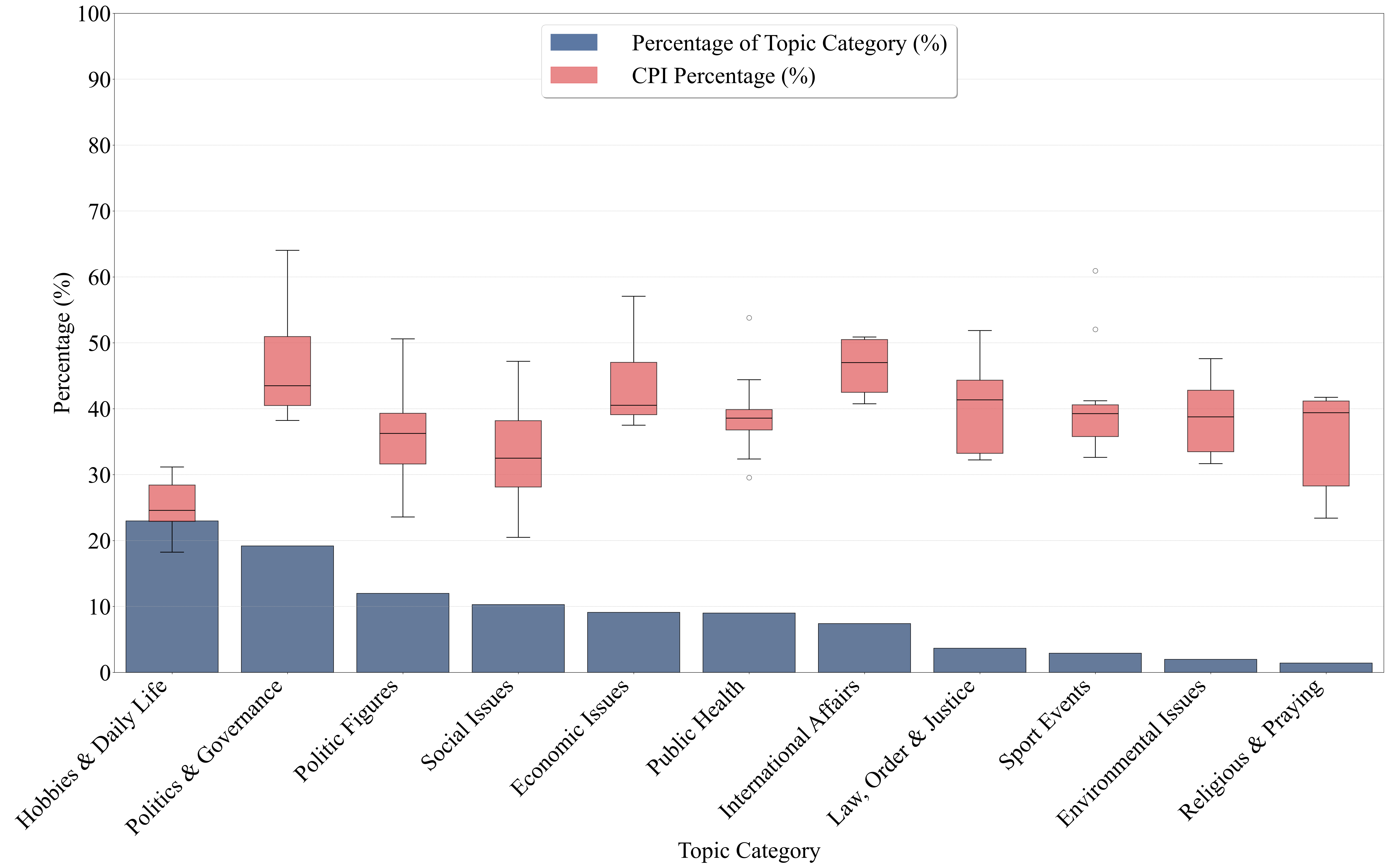}
        \caption{\hl{All}}
        \label{fig:topics-cpi-distro-all}
        \end{subfigure}
        \hfill
        \centering
        \begin{subfigure}[b]{\linewidth}
        \centering
        \includegraphics[width=0.8\linewidth]{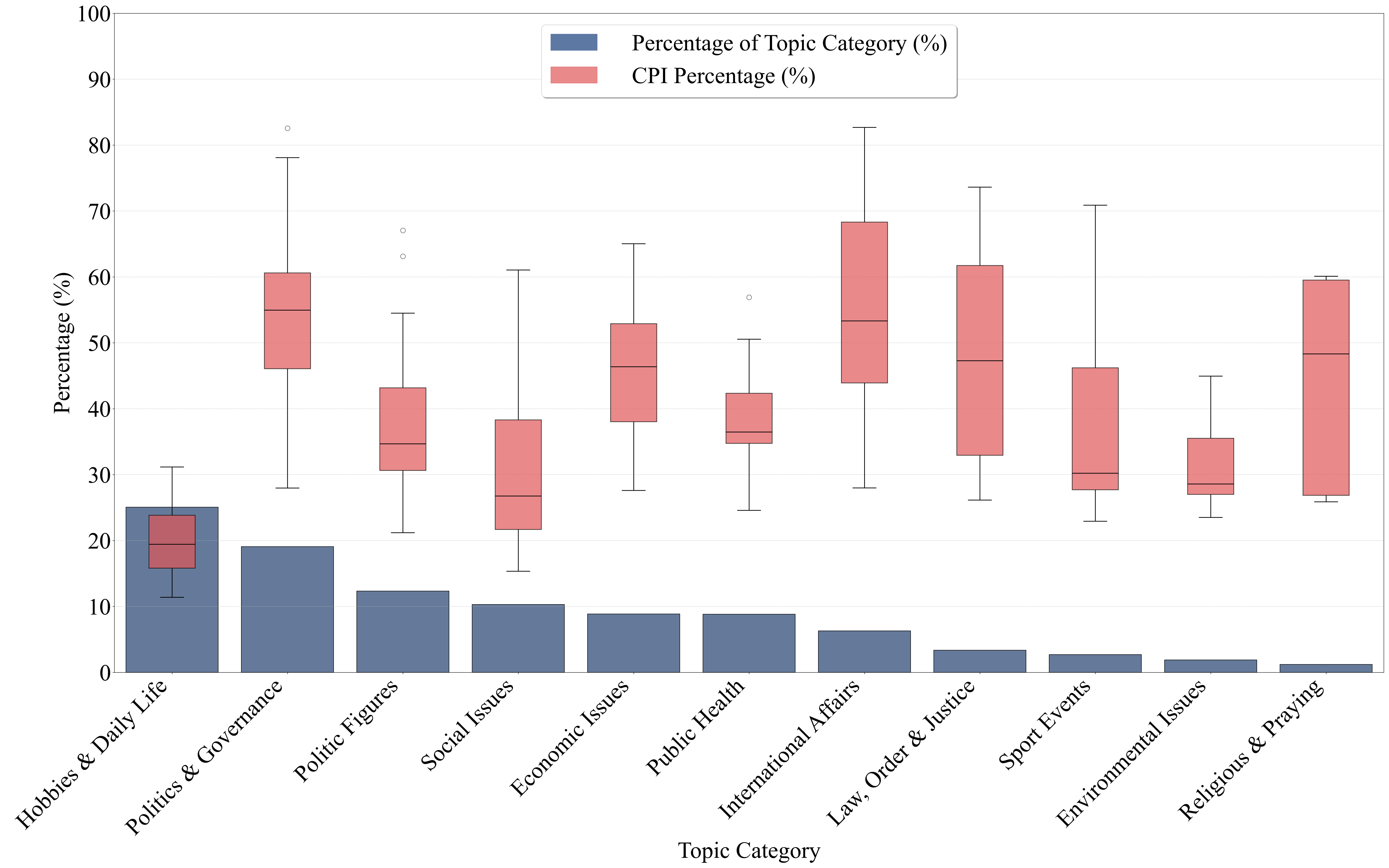}
        \caption{\hl{Democrats}}
        \label{fig:topics-cpi-distro-democrats}
        \end{subfigure}
        \hfill
        \begin{subfigure}[b]{\linewidth}
        \centering
        \includegraphics[width=0.8\linewidth]{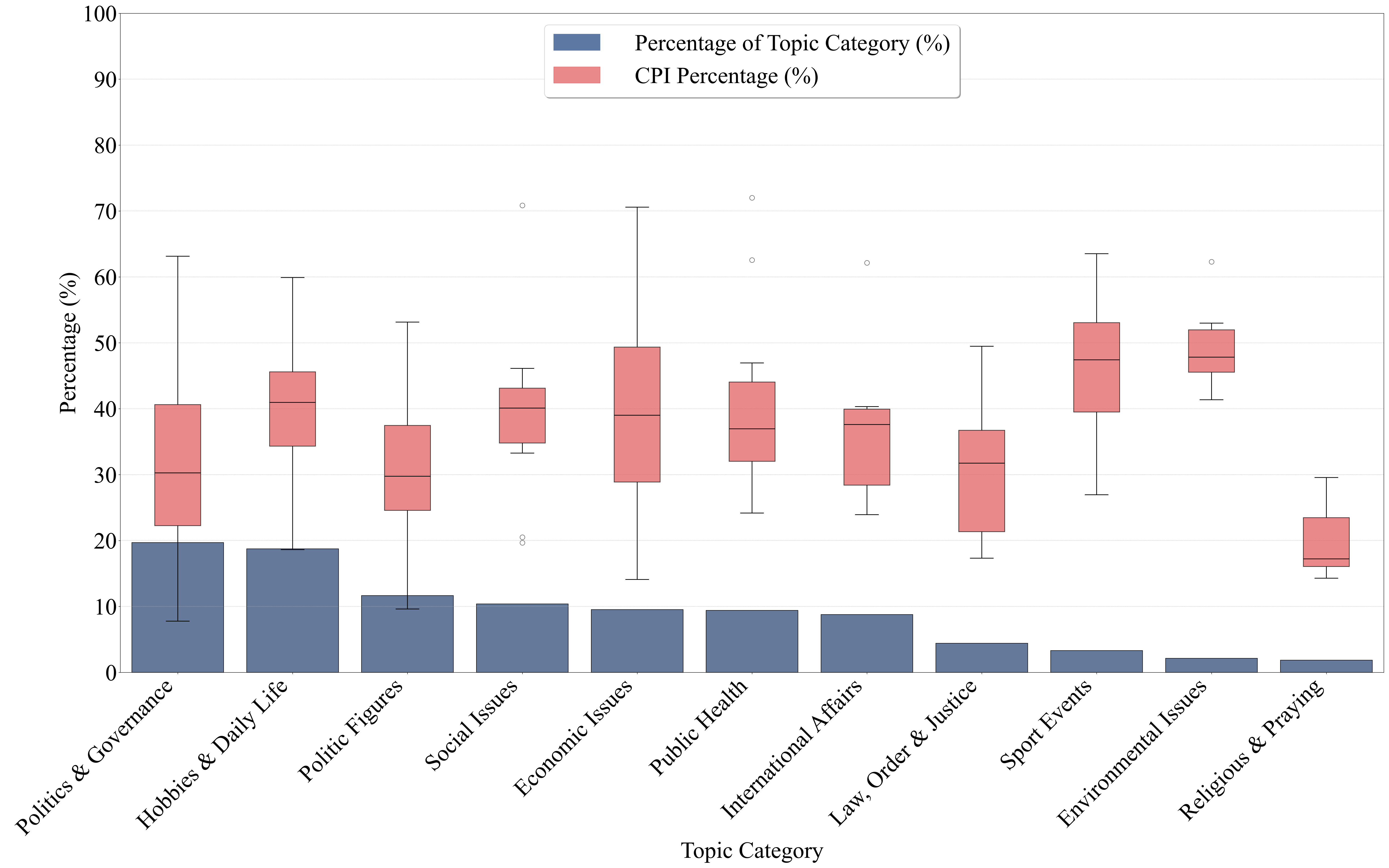}
        \caption{\hl{Republicans}}
        \label{fig:topics-cpi-distro-republicans} 
        \end{subfigure}
        \hfill
        \caption{The distribution of topic categories in the dataset with the box-and-whisker plot of their CPI percentages \hl{on different topics for the entire dataset and both camps}.}
        \label{fig:topics-cpi-distro} 
\end{figure}

We compute the percentage of replies to root tweets in a specific category relative to all categories. We also compute the percentage of CPIs in each category. Figure~\ref{fig:topics-cpi-distro} presents the results. The replies to tweets with the topics under \textit{Hobbies \& Daily Life}, and \textit{Politics \& Governance} categories emerge as the most substantial, making up nearly half of the data \hl{for both camps}. In contrast, \textit{Law, Order \& Justice}, \textit{Sport Events} \textit{Environmental Issues}, and \textit{Religious \& Praying} occupy smaller portions of the dataset as they are rather niche. The popularity of categories remains consistent across both groups, except for a notable difference: \textit{Hobbies \& Daily Life} is the most popular among Democrats, whereas \textit{Politics \& Governance} takes the lead among Republicans.

The topics under \textit{Politics \& Governance} and \textit{Politic Figures} generate the highest levels of CPI, suggesting that they are particularly effective at breaking through echo chambers. Within these politically charged categories, there is considerable variance in CPI percentages, reflecting the context-dependent nature of how these topics are discussed and debated. For instance, topics such as ``Impeachment Proceedings and Senate Actions'' and ``Election Strategies'' exhibit a high degree of CPI (53, 60\%), highlighting their contentious nature and the widespread debate across ideological divides.

The topics of public interest such as ``Political Approval'', ``Sporting Events'', and ``Economic Stimulus'' exhibit significant cross-partisan engagement. These topics span various domains, including governance, sports, and economic issues, attracting users from diverse ideological backgrounds due to their broad relevance. Discussions related to \textit{Sport Events} like "Sporting Events and National Interests" show notable CPI with percentages above 55\%. In contrast, more routine discussions in \textit{Public Health}, such as ``Health Guidelines and Mask Usage'' and ``Hygiene Practices'', display lower CPI, with percentages between 27\% and 36\%. 

A contrast between camps is observed in the ``Hobbies \& Daily Life'' and ``Politics \& Governance.'' While both groups exhibit a significant proportion of discussions in these categories, Republicans demonstrate a higher CPI percentage in the ``Hobbies \& Daily Life.'' In contrast, Democrats show a slightly higher CPI percentage in the ``Politics \& Governance.'' The category ``Sport Events'' stands out due to its consistently high CPI percentages for both groups.

Conversely, topics such as ``Religious \& Praying'' and ``Environmental Issues'' are less represented in the dataset, exhibiting minimal proportions for both groups. However, Democrats show a slightly higher CPI percentage in ``Religious \& Praying,'' suggesting a marginally greater propensity for cross-partisan discourse on this topic. Meanwhile, discussions on ``International Affairs'' appear consistently across both groups but differ in CPI percentages. Democrats display a higher median CPI percentage and a broader range of cross-partisan engagement in this category.

Topics under the categories \textit{Hobbies \& Daily Life} and \textit{Religious \& Praying} exhibit considerably lower CPI percentages, indicating that they are more likely to be confined within echo chambers. For instance, topics such as ``\textit{Musical Themes}'', ``\textit{Reading Habits}'', ``\textit{Popular Food Discussions}'', ``\textit{Community Support and Prayers}'', ``\textit{Religious Cultural Commentary}'', and `\textit{`Religious Beliefs}'' show minimal cross-partisan engagement, with CPI percentages between 23\% and 30\%, which suggests that users participating in these discussions are more insular, predominantly interacting with like-minded individuals. However, there are exceptions, particularly within the sports category, where discussions around Golf and Football topics show higher CPI percentages, ranging from 51\% to 55\%. 

The findings suggest that while politically and socially significant topics tend to disrupt echo chamber dynamics, more personal or neutral topics may reinforce them, leading to isolated discussions within homogeneous groups. While echo chambers are indeed a concern, particularly around less contentious topics, the data reveals that not all discussions are confined to such insular spaces. Political and controversial subjects seem to break through these barriers, introducing users to diverse perspectives, a beneficial trend for democratic processes, and informed civic engagement.

\subsection{Topics vs. Stance \& Toxicity}
\begin{figure*}
    \centering
    \includegraphics[width=0.9\linewidth]{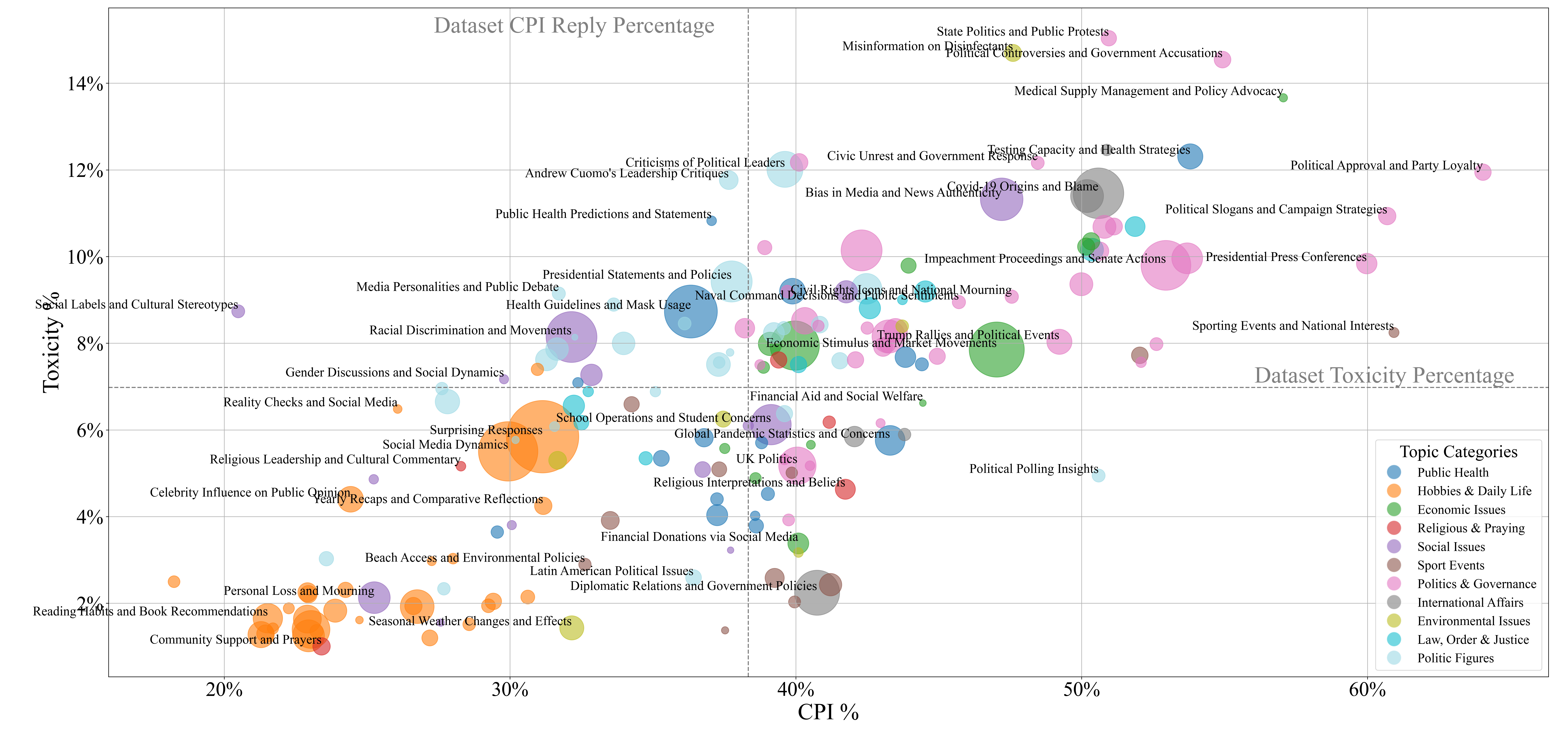}
    \caption{The distribution of the topics with axes representing the CPI percentage and the toxicity percentage.}
    \label{fig:topics-cpi-toxicity-distro}
\end{figure*}
The variance in CPI and toxicity percentages across different topics, as illustrated in Figure~\ref{fig:topics-cpi-toxicity-distro}, offers valuable insights into the dynamics of online discourse. Generally, CPI percentages reflect a topic's political or social sensitivity, with those rooted in societal values and governance tending to exhibit higher CPI rates, sparking greater cross-partisan discussions. A Pearson correlation of 0.76 between topic CPI and toxic content percentages emphasizes the strong positive relationship between these two dimensions. 

Topics in the chart's top-left quadrant, such as ``\textit{Social Labels and Cultural Stereotypes},'' ``\textit{Gender Discussions},'' and ``\textit{Racial Discrimination},'' represent highly toxic, low-CPI subjects. These are problematic as they often fuel internal group animosity and deepen polarization. In contrast, topics in the bottom-right quadrant, such as ``\textit{School Operations and Student Concerns},'' ``\textit{Financial Aid and Social Welfare},'' ``\textit{NFL Draft Strategies},'' and ``\textit{Health Insurance Policies},'' are characterized by high-CPI and lower toxicity, which offer opportunities for bridging societal divides, acting as contact points where cross-partisan dialogue can flourish.

 \begin{figure}[!htb]
        \centering
        \includegraphics[width=\linewidth]{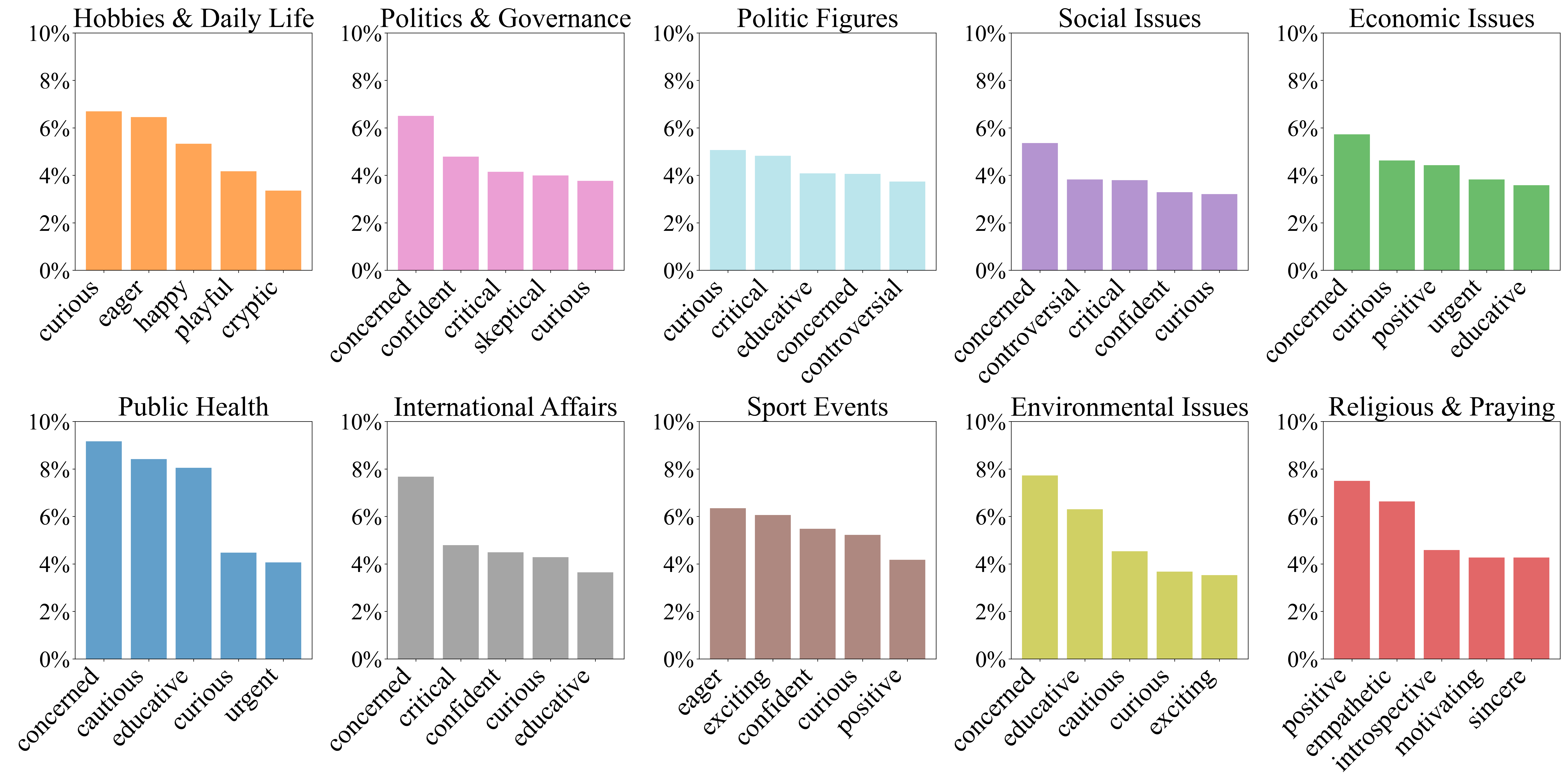}
        \caption{Top 5 reply stance annotation percentages under each topic for CPIs.}
        \label{fig:topic-reply-stance-plot}    
\end{figure}

Figure~\ref{fig:topic-reply-stance-plot} presents the top five stance percentages used in CPIs across major topics, offering insights into the emotional and attitudinal tone of the replies. The distribution of stances varies significantly depending on the topic, highlighting the diversity of cross-partisan communications. For example, in non-political domains like \textit{Hobbies \& Daily Life} and \textit{Sport Events}, positive and playful stances such as ``\textit{eager},'' ``\textit{happy},'' and ``\textit{exciting}'' are prominent, suggesting that these areas foster more open and lighthearted dialogue. Conversely, topics related to \textit{Politics \& Governance} and \textit{Public Health} are dominated by more serious and critical stances, including ``\textit{concerned},'' ``\textit{confident},'' and ``\textit{cautious},'' reflecting these discussions' contentious and urgent nature. ``\textit{Religious \& Praying}'' demonstrates a higher prevalence of empathetic and introspective stances, indicating a more emotionally driven discourse. The figure also reveals that certain topics, such as \textit{Economic Issues} and \textit{International Affairs,} exhibit a blend of concerned and curious stances, pointing to an underlying tension between critique and a willingness to understand the opposing view. The variation in stance distribution across topics suggests that the emotional tenor of CPIs is context-dependent, with some domains providing opportunities for constructive engagement while others remain more adversarial or cautious. Understanding these stance patterns is critical for fostering healthier cross-partisan dialogue in polarized environments.

\section{RQ3: Content Analysis}
\label{sec:content-analysis}

Lastly, we investigate the content characteristics that are more associated with CPIs as part of our efforts to address RQ3. 
For this, we leverage the LLM annotations (root sentiments and reply stances) generated in Section~\ref{sec:method}.

\subsection{Stance Contrast, PI vs. CPI}
\label{sec:rq3:stance}

Our initial phase of content analysis is to investigate how users of different parties differentiate in terms of stance when interacting with in-group (PI) and out-group (CPI) users. This can potentially shed light on the productivity of CPIs across the two parties.

We compute the frequency differences of each AI-annotated stances of the replies across the PI and CPI of every party (i.e. D\textrightarrow D vs. D\textrightarrow R, and R\textrightarrow R vs. R\textrightarrow D) and normalize them by the average of the annotation frequency between them. Then, we visualize the contrast for every annotation in a pyramid bar chart. Let $f^{G_{right}}_i$ denote the frequency of the AI-provided annotation $i$ in the group on the right side of the bar, the x-axis of the chart is calculated as in Equation~\ref{eq:x_barchart}:

\begin{equation}
X_i = \frac{f^{G_{right}}_i - f^{G_{left}}_i}{(f^{G_{right}}_i + f^{G_{left}}_i)\times0.5}
\label{eq:x_barchart}
\end{equation}

A positive (negative) value indicates a higher frequency in the right (left) group.

\begin{figure}[!htb]
        \centering
        \begin{subfigure}[b]{\linewidth}
        \centering
        \includegraphics[width=0.9\linewidth]{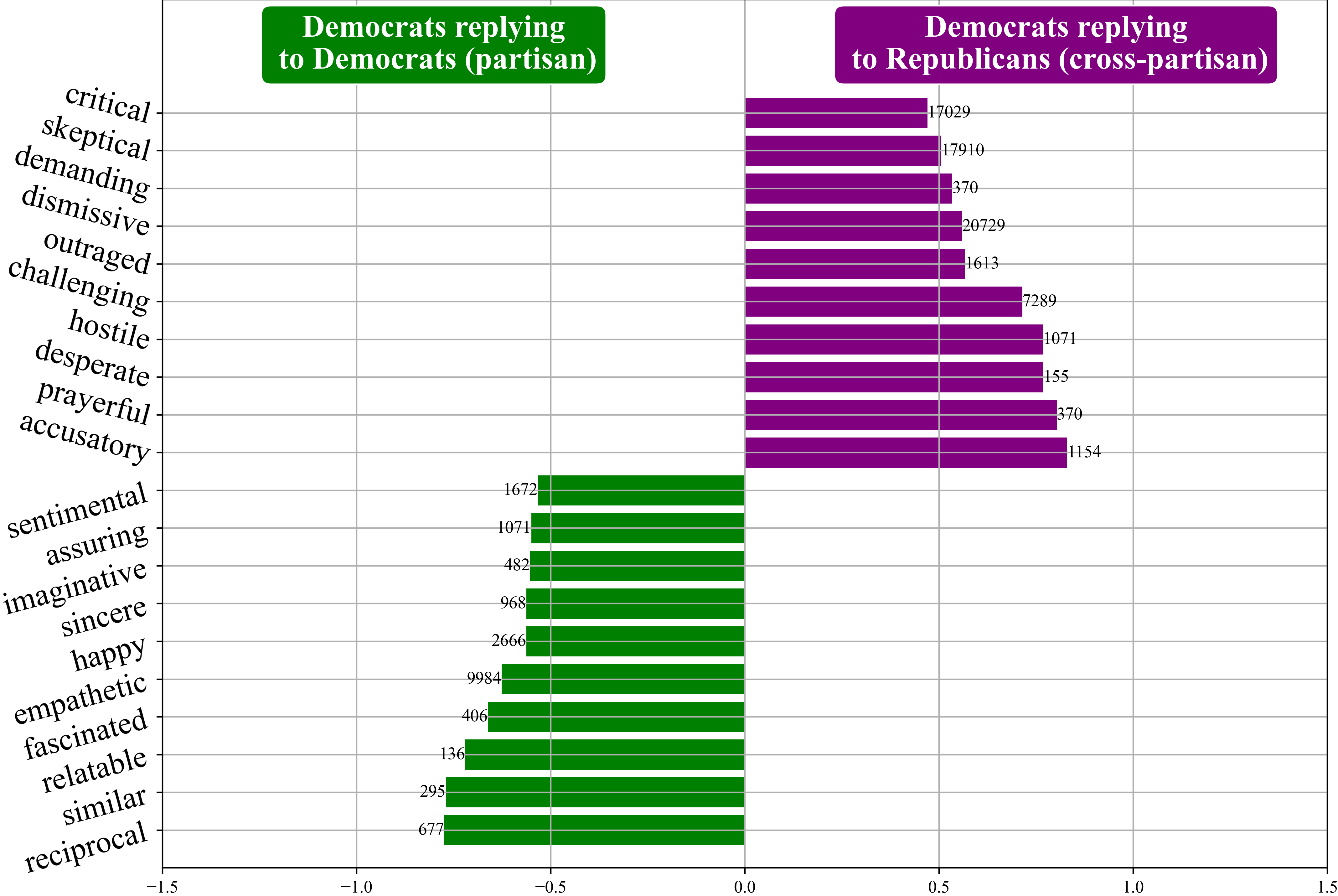}
        \caption{$D \rightarrow D$ \textbf{vs.} $D \rightarrow R $}
        \label{fig:D_D_vs_R_D}
        \end{subfigure}
        \hfill
        \begin{subfigure}[b]{\linewidth}
        \centering
        \includegraphics[width=0.9\linewidth]{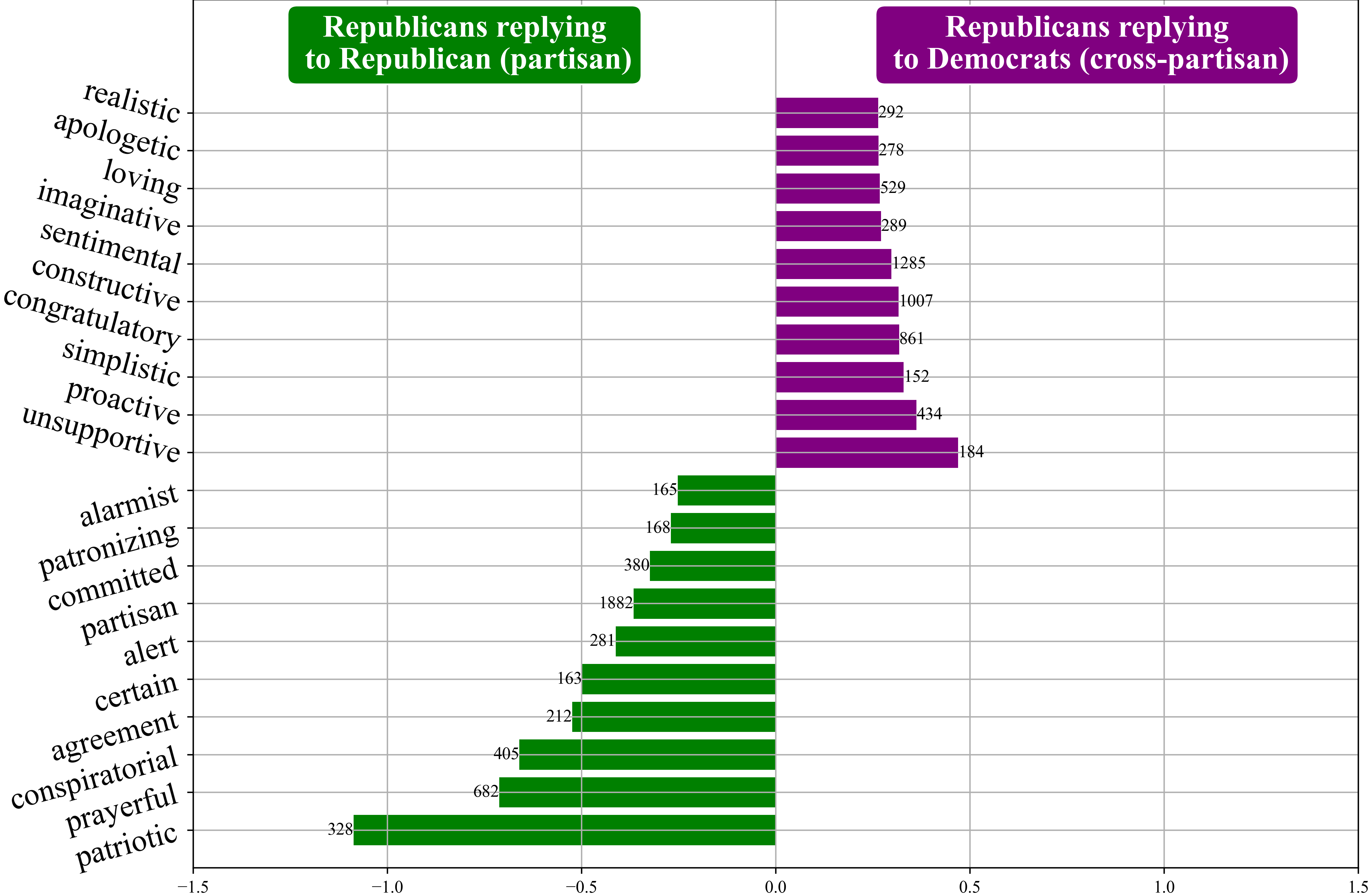}
        \caption{$R \rightarrow R$ \textbf{vs.} $R \rightarrow D$}
        \label{fig:R_R_vs_D_R}
        \end{subfigure}
        \hfill
        \caption{Stance-wise differences of partisan \textbf{vs.} cross-partisan replies across parties. The bar labels indicate the overall frequency of the annotation. For samples of tweets for each annotation (stance), see Table~\ref{tab:samples-stance} in Appendix~\ref{sec:appendix}. 
        }
        \label{fig:barcharts-PI-CPI}    
\end{figure}

Figure~\ref{fig:D_D_vs_R_D} presents the word frequency differences between Democrat-to-Democrat (D\textrightarrow D) and Democrat-to-Republican (D\textrightarrow R) interactions. Notably, D\textrightarrow D interactions are characterized by words associated with empathy, positive sentiment, and agreement, such as ``\textit{happy}'', ``\textit{reciprocal}'', and ``\textit{empathetic}''. Conversely, D\textrightarrow R interactions exhibit words indicative of conflict, like ``\textit{accusatory}'', ``\textit{dismissive}'', ``\textit{hostile}'', and ``\textit{critical}''.

Unlike Democrats, we do not observe a strong contrast between R$\rightarrow$R and R$\rightarrow$D interactions (Figure~\ref{fig:R_R_vs_D_R}). Except for ``\textit{agreement}'' vs. ``\textit{unsupportive}'', Republican replies show more consistent tone across partisan and cross-partisan contexts. Some positive stances, like ``\textit{loving}'' and ``\textit{congratulatory}'', appear more often in R$\rightarrow$D replies, often in personal contexts such as births or marriages (Table~\ref{tab:samples-stance}). This aligns with our earlier finding that Republicans tend to reply to less-followed users (Table~\ref{tab:logit_results}), suggesting a lower distinction between in- and out-group engagement, particularly in personal matters. Furthermore, the stance contrast bars in Figure~\ref{fig:D_D_vs_R_D} are visibly larger than in Figure~\ref{fig:R_R_vs_D_R}, indicating Democrats vary their tone more between PIs and CPIs. While Democrats are more likely to engage in CPIs~\cite{barbera2015birds}, their cross-partisan replies are also more likely to be negative or confrontational compared to their intra-party ones—a pattern less evident among Republicans.

\subsection{Root Sentiment vs. Reply Stance \& Toxicity}
\label{sec:rq3:rootvsreply}

\begin{figure}[ht]
        \centering
        \includegraphics[width=0.8\linewidth]{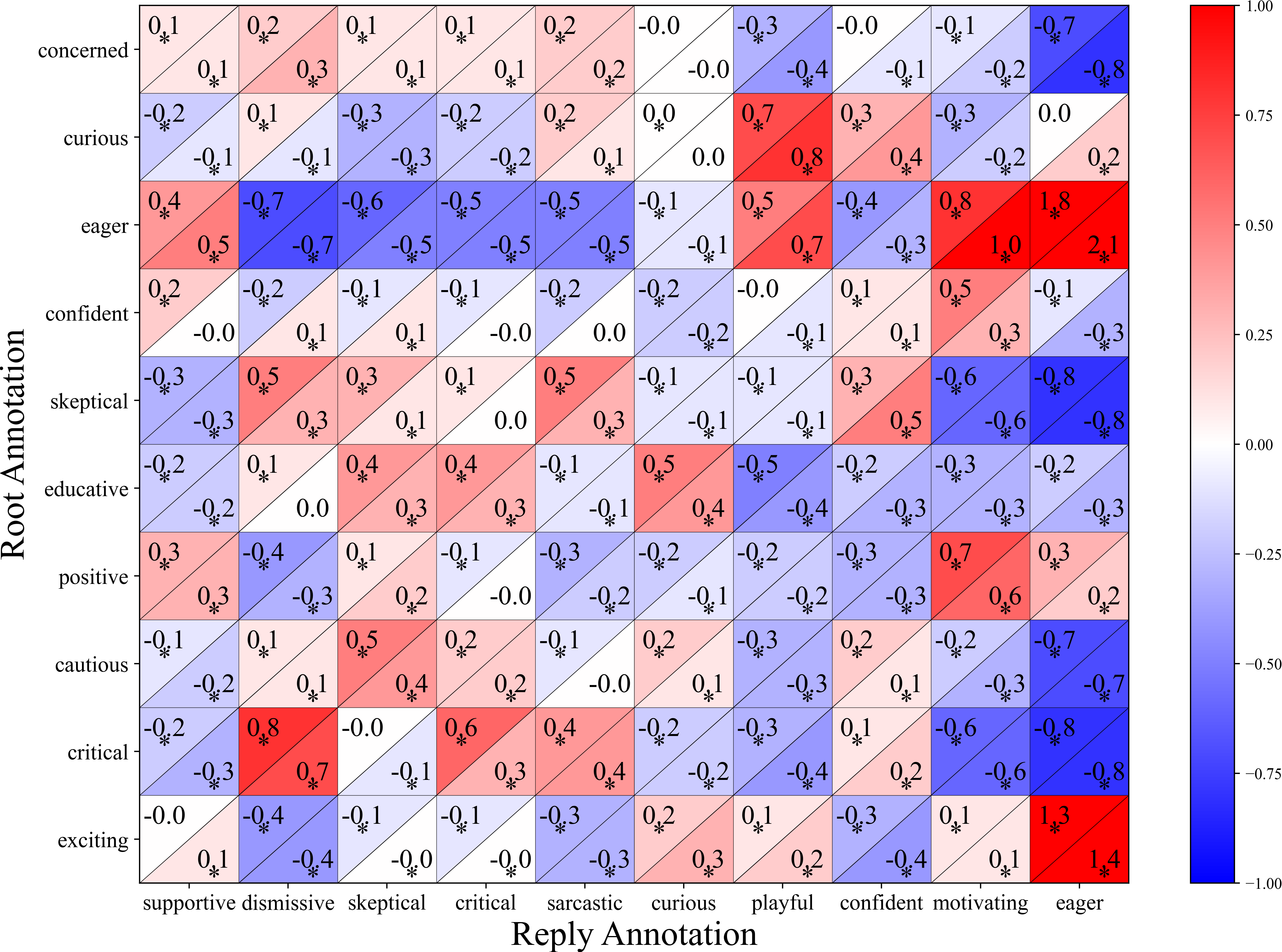}
        \caption{Chi-test statistics $\chi = \frac{Observed - Expected}{Expected}$ for co-occurrences of sentiments in root tweets and stances in replies (Top-Left: PI, Bottom-Right: CPI). Starred cells indicate p-values below 0.05.}
        \label{fig:root-reply-contingency}    
\end{figure}

Our next content-based analysis focuses on identifying the underlying sentiments that are more likely to foster certain stances in replies.
We employ the top 10 most frequent adjectives from the root tweets' sentiments and stances of the replies to create contingency matrices for both PI and CPI. Using Chi-Squared test $\chi^2 = \sum \frac{(Observed - Expected)^2}{Expected}$ we compute the association between each sentiment$\leftrightarrow$stance pair. 
Figure~\ref{fig:root-reply-contingency} visualizes the sentiment$\leftrightarrow$stance relationship by a triangular heatmap, where each cell was divided into two: the top-left triangle shows the values for the partisan interactions, while the bottom-right triangle represents cross-partisan interactions. The color scale, ranging from blue (-1) to red (+1) with white at 0, reflects the strength of the observed versus expected values. Therefore, a positive (negative) value in each cell indicates a positive (negative) association between the corresponding root sentiment and reply stance. Each cell on Figure~\ref{fig:root-reply-contingency} shows $\frac{Observed - Expected}{Expected}$ to indicate positive and negative associations and the p-values. The p-values are computed after taking the square of the nominator as in the original setting.

The heatmap reveals that certain root sentiments are more likely to encourage similar sentiments/stances in the replies. For instance, we observe that ``\textit{eager}'' sentiment in the root tweets often lead to ``\textit{eager}'' and ``\textit{motivating}'' stances in replies, with values highly above 0, indicating a strong positive correlation. Similarly, ``\textit{Positive}'' sentiments generally elicit more ``\textit{motivating}'' responses, while ``\textit{critical}'' sentiments lead to stances like ``\textit{dismissive}'', ``\textit{critical}'', and ``\textit{sarcastic}'', all with values greater than 0. In contrast, ``\textit{skeptical}'' and ``\textit{critical}'' sentiments discourages ``\textit{eager}'' and ``\textit{motivating}'' stances in replies, with values strongly below 0.

We also observe that these patterns are relatively consistent across both partisan and cross-partisan interactions in the heatmap. To further validate this, we extend the scope of the contingency matrix to the 50 most frequent sentiments and stances (a contingency matrix with $2500\times2$ cells) and detect a Pearson correlation of 95\% between the values calculated for PIs and the values calculated for the CPIs. This may suggest that root sentiment is not a very strong factor for fostering CPIs compared to users and topics.

\begin{figure}[ht]
        \centering
        \includegraphics[width=\linewidth]{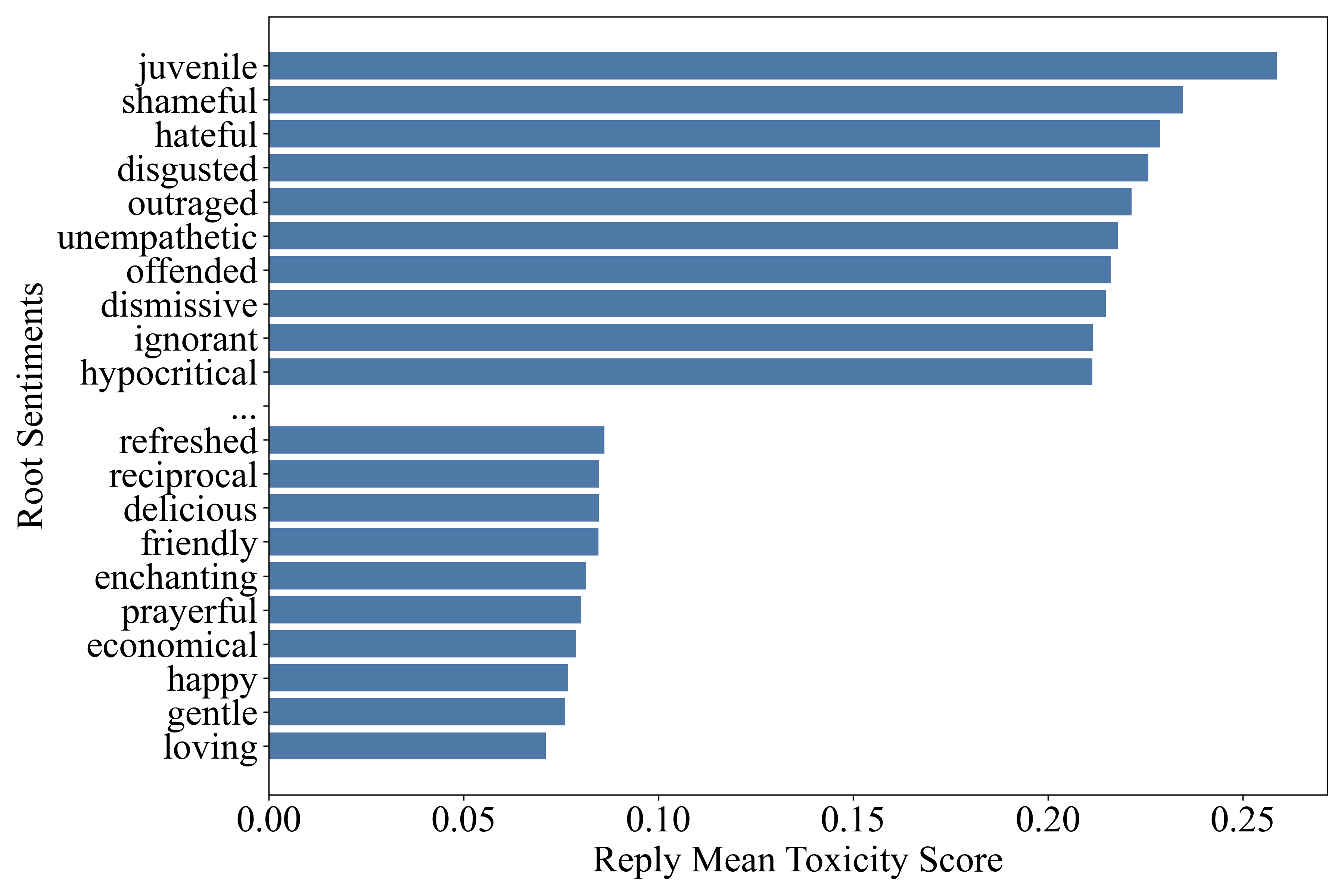}
        \caption{Top 10 root sentiments that attract the highest and lowest toxicity score replies.}
        \label{fig:root-sentiment-reply-toxicity}    
\end{figure}

In addition to stances, we also investigate the toxicity rates provoked in the replies by each of the root sentiments.
Figure~\ref{fig:root-sentiment-reply-toxicity} illustrates the top 10 root sentiments that attract the highest and lowest toxicity score replies in cross-partisan interactions. The mean toxicity score quantifies the level of hostility or aggressive language in replies for the given root sentiments that provoke these responses. The data shows that negative sentiments such as ``\textit{juvenile},'' ``\textit{shameful},'' ``\textit{hateful},'' and ``\textit{disgusted}'' are associated with the highest mean toxicity scores, ranging between 0.18 and 0.25. These sentiments, often judgmental or derogatory, may incite defensive or antagonistic replies, particularly in cross-partisan settings where opposing ideological stances intensify interpersonal conflict. In contrast, more positive or neutral sentiments like ``\textit{loving},'' ``\textit{gentle},'' ``\textit{happy},'' and ``\textit{friendly}'' correspond to significantly lower toxicity scores, ranging from 0.04 to 0.08. These sentiments indicate harmonious or empathetic discourse, fostering more civil and respectful replies. The stark contrast between the sentiments that provoke the most and least toxic responses suggests that the nature of root sentiments plays a pivotal role in shaping the tone of online discussions. In CPIs, where political and ideological tensions are inherent, sentiment-driven language influences the trajectory of dialogue, with more aggressive language fueling hostility, and gentler or more neutral language encouraging more constructive exchanges. Therefore, strategically using language emphasizing empathy or neutrality may mitigate the toxicity in polarized conversations. 

\vspace{-1.em}

\section{Discussion}
\label{sec:discussion}

We discuss the takeaways of our analysis and its limitations. 

\subsection{Takeaways}

Our work provides valuable new insights into online CPIs from a multi-fold perspective, i.e.: user, topic, and content. 
In particular, we find common characteristics of users engaging in CPIs, analyzing the discourse in terms of topics, and content; where the content breaks down to the analysis of \textit{sentiments} of the roots, \textit{stances} of the replies, and \textit{toxicity}.
When we put these three dimensions in context, we see how social media exchanges influence societal polarization and discourse quality from a complementary perspective when compared to prior work. In particular, we derive the following novel insights and discuss the implications they have.

\noindent\textbf{The Who:}
Those who participate in CPI show {\em distinct patterns based on their political orientation}. 
Users with extreme political views, especially on the far-right, tend to use more toxic language. 
Interestingly, the center-right replies with less \textit{toxicity} than the center-left, yet the \textit{extreme toxicity} category is still higher in Republicans. 
Although Democrats are proven to be more likely to engage in CPI than Republicans~\cite{barbera2015birds}, their participation often includes more negative and non-constructive stances, such as being ``\textit{critical}'', ``\textit{accusatory}'', or ``\textit{hostile}''. 
In contrast, Republicans show less differentiation in stance between interactions within their own group (PI) and cross-partisan replies (CPI), \hl{with instances of even positive CPI stances (``\textit{loving}'' and ``\textit{congratulatory}'')}, suggesting a more consistent approach to discourse regardless of political affiliation.

Moreover, we detect an association between the CPI rate of a user and their popularity:
{\em Popular users receive more attention, leading to higher CPIs, while smaller accounts mainly engage with the opposite group.} One potential explanation for this could be that since the less followed accounts are viewed less, they may feel more secure to engage in a controversial debate with the opposite party. Alternatively, they may be fake or new \& small accounts craving for increasing their impressions and followers.

\noindent\textbf{The What:}
{\em CPIs are significantly influenced by the nature of the topics being discussed, with politically charged events acting as strong catalysts.}
However, not all topics foster constructive dialogue. 
Topics such as gender, race, and discrimination tend to foster less CPI but higher toxicity, indicating more adversarial and polarized exchanges. 
Conversely, discussions around financial aid and social welfare demonstrate a healthier balance, with more CPI and lower toxicity, suggesting these areas offer opportunities for more constructive and cooperative cross-party engagement.
Identifying topics with low toxicity and high CPI can help highlight potential opportunities for reducing polarization, while high-toxicity, low-CPI topics represent areas of concern. 
{\em Our results show the distribution of CPI and toxicity across different discussion themes, highlighting potential opportunities for reducing polarization and identifying areas of concern.}

\noindent\textbf{The How:} {\em To foster a healthy online dialogue, how you talk matters.} 
Users who talk in a friendly way, receive less toxic or critical replies. For instance, in Figure~\ref{fig:root-reply-contingency}, ``\textit{eager}'' and ``\textit{positive}'' sentiments in the root tweets associate with ``\textit{motivating}'' and ``\textit{eager}'' stances in the replies and correlate negatively with ``\textit{dismissive}'' stance. On the other hand, negative sentiments in root tweets are more likely to provoke toxic (see Figure~\ref{fig:root-sentiment-reply-toxicity}) or negative stance in the replies (e.g. ``\textit{dismissive}'' or ``\textit{critical}''). Interestingly, this phenomenon is independent of whether the interaction is a CPI or a PI as we observed a 95\% correlation for the values in Figure~\ref{fig:root-reply-contingency}.
{\em Understanding this dynamic is paramount to reducing the widespread polarization in online spaces and disrupting the idea that toxicity is inherent to certain online communities.}

\subsection{Limitations}
\label{sec:discussion:limitation}

\noindent{\bf Continuous CPIs}: The CPIs between users close to the political center, and the CPIs between extremes may be inherently different. We initially experimented with a continuous CPI value by multiplying the political orientation score of the poster and the replier. However, we did not notice a drastic change in the results of our initial experiments. We plan to address a deeper analysis as part of our future work.

\noindent{\bf Reply chain and media}: To simplify the analysis, we limit it to root tweets and their direct replies, and discard the replies of replies. 
Our content analysis is limited to the textual content present in the tweets. The analysis of the links, images, and videos in the tweets falls beyond our scope.

\noindent{\bf LLM's annotation}: We used a heuristic approach with Mistral-7B-Instruct-v0.2 for stance/sentiment annotation. While effective overall, some replies were misclassified due to overlap between stance and sentiment. 
Without setting a limit for the number of adjectives, the LLM generated sentences rather than words as in Table~\ref{tab:prompt-sample}. Moreover, in some cases, the LLM did not differentiate the concept of \textit{stance} and \textit{sentiment} when asked to annotate the \textit{stances} of the replies (e.g. the word ``loving'' in Figure~\ref{fig:barcharts-PI-CPI} is more of a \textit{sentiment} than a \textit{stance}). 
However, we find such confusions to be rare, and we observe that they do not hinder the objective of our analysis. 
This is because understanding any salient semantic quality in the content of replies is insightful for us; whether it is \textit{stance} or \textit{sentiment}. 
Experiments with other LLMs and configurations are left to the future work.

\noindent{\bf Platform}: \hl{Our study focuses on X (formerly Twitter), which has its own unique dynamics. These dynamics may influence the interaction patterns and content characteristics observed in our analysis. Expanding this research to other social media platforms, such as Reddit, YouTube, Instagram, and others, would provide a broader understanding of cross-platform user behavior and interactions. Such platforms offer different interaction paradigms, media-sharing mechanisms, and community structures that could yield additional insights and validate the generalizability of our findings.}

\noindent{\bf Context}: \hl{The data collection period was chosen to be 2020 as this year has seen massive debates both on political (riots, elections) and non-political debates (the pandemic). However, the presence of a Republican government in the U.S. may have influenced the results, which could differ under a Democrat government. Additionally, the results may have differed during a period with less politically charged events, such as in 2021. Extending the study over a longer period could help capture evolving patterns.  We acknowledge these limitations and leave extending the study to a longer period for future work due to current restrictions with collecting Twitter data. Finally, as the dataset was limited to the U.S. context, the findings may not be a global context.}

\section{Ethical Impact}
\label{sec:conclusion}

In this work, we collect and analyze the posts and replies among users and their profile attributes. We collect this data using Twitter's Academic API before May 2023. We mitigate the potential ethical risks of analyzing user data by limiting our focus to public accounts and analyzing aggregate data. That is, we do not examine and report individual profiles. To further protect user privacy, we anonymize the user data before sharing our dataset. Our regression models do not use any personally identifiable information. Our study aims to bridge divides among the public by understanding the dynamics of non-toxic cross-partisan interactions. It should not be misused to create or amplify content that would provoke toxic content within echo chambers or manipulate the public discourse in any other way.

\section*{Acknowledgments}

Guillermo Suarez-Tangil has been appointed as 2019 Ramon y Cajal fellow (RYC-2020-029401-I) funded by MICIU/AEI/\-10.13039/\-5011\-000\-11033 and ESF Investing in your Future. 

\bibliography{references}

\section*{Ethics Checklist}

\begin{enumerate}

\item For most authors...
\begin{enumerate}
    \item  Would answering this research question advance science without violating social contracts, such as violating privacy norms, perpetuating unfair profiling, exacerbating the socio-economic divide, or implying disrespect to societies or cultures?
    \answerYes{Yes}
  \item Do your main claims in the abstract and introduction accurately reflect the paper's contributions and scope?
    \answerYes{Yes}
   \item Do you clarify how the proposed methodological approach is appropriate for the claims made? 
    \answerYes{Yes}
   \item Do you clarify what are possible artifacts in the data used, given population-specific distributions?
    \answerYes{Yes}
  \item Did you describe the limitations of your work?
    \answerYes{Yes, check Section~\ref{sec:discussion:limitation}} 
  \item Did you discuss any potential negative societal impacts of your work?
    \answerNA{NA}
      \item Did you discuss any potential misuse of your work?
    \answerNA{NA}
    \item Did you describe steps taken to prevent or mitigate potential negative outcomes of the research, such as data and model documentation, data anonymization, responsible release, access control, and the reproducibility of findings?
    \answerYes{Yes, the data we used was open-sourced and our analysis does not reveal the identity of the users. Also we share our labels on GitHub for reproducibility.}
  \item Have you read the ethics review guidelines and ensured that your paper conforms to them?
    \answerYes{Yes}
\end{enumerate}

\item Additionally, if your study involves hypotheses testing...
\begin{enumerate}
  \item Did you clearly state the assumptions underlying all theoretical results?
    \answerYes{Yes}
  \item Have you provided justifications for all theoretical results?
    \answerYes{Yes}
  \item Did you discuss competing hypotheses or theories that might challenge or complement your theoretical results?
    \answerYes{Yes}
  \item Have you considered alternative mechanisms or explanations that might account for the same outcomes observed in your study?
    \answerYes{Yes}
  \item Did you address potential biases or limitations in your theoretical framework?
    \answerYes{Yes}
  \item Have you related your theoretical results to the existing literature in social science?
    \answerYes{Yes, in Section~\ref{sec:discussion}, we make references to previous work and compare our findings.}
  \item Did you discuss the implications of your theoretical results for policy, practice, or further research in the social science domain?
    \answerYes{Yes}
\end{enumerate}

\item Additionally, if you are including theoretical proofs...
\begin{enumerate}
  \item Did you state the full set of assumptions of all theoretical results?
    \answerNA{NA}
	\item Did you include complete proofs of all theoretical results?
    \answerNA{NA}
\end{enumerate}

\item Additionally, if you ran machine learning experiments...
\begin{enumerate}
  \item Did you include the code, data, and instructions needed to reproduce the main experimental results (either in the supplemental material or as a URL)?
    \answerYes{Yes, the code is fully shared online.}
  \item Did you specify all the training details (e.g., data splits, hyperparameters, how they were chosen)?
    \answerYes{Yes}
     \item Did you report error bars (e.g., with respect to the random seed after running experiments multiple times)?
    \answerNA{NA. Our machine learning experiment was done for generating text where the concept of accuracy and error-bar does not apply.}
	\item Did you include the total amount of compute and the type of resources used (e.g., type of GPUs, internal cluster, or cloud provider)?
    \answerYes{Yes, see Section~\ref{sec:method}}
     \item Do you justify how the proposed evaluation is sufficient and appropriate to the claims made? 
    \answerYes{Yes}
     \item Do you discuss what is ``the cost`` of misclassification and fault (in)tolerance?
    \answerYes{Yes}
  
\end{enumerate}

\item Additionally, if you are using existing assets (e.g., code, data, models) or curating/releasing new assets, \textbf{without compromising anonymity}...
\begin{enumerate}
  \item If your work uses existing assets, did you cite the creators?
    \answerYes{Yes}
  \item Did you mention the license of the assets?
    \answerYes{Yes}
  \item Did you include any new assets in the supplemental material or as a URL?
    \answerYes{Yes}
  \item Did you discuss whether and how consent was obtained from people whose data you're using/curating?
    \answerNA{NA}
  \item Did you discuss whether the data you are using/curating contains personally identifiable information or offensive content?
    \answerYes{Yes}
\item If you are curating or releasing new datasets, did you discuss how you intend to make your datasets FAIR (see \citet{fair})?
\answerNA{NA}
\item If you are curating or releasing new datasets, did you create a Datasheet for the Dataset (see \citet{gebru2021datasheets})? 
\answerNA{NA}
\end{enumerate}

\item Additionally, if you used crowdsourcing or conducted research with human subjects, \textbf{without compromising anonymity}...
\begin{enumerate}
  \item Did you include the full text of instructions given to participants and screenshots?
    \answerNA{NA}
  \item Did you describe any potential participant risks, with mentions of Institutional Review Board (IRB) approvals?
    \answerNA{NA}
  \item Did you include the estimated hourly wage paid to participants and the total amount spent on participant compensation?
    \answerNA{NA}
   \item Did you discuss how data is stored, shared, and deidentified?
   \answerNA{NA}
\end{enumerate}

\end{enumerate}
\appendix
\section{Appendix}
\label{sec:appendix}

\begin{table*}[ht]
\centering
\vspace{-8pt}
\begin{tabular}{p{2cm} | p{14cm}}

\hline
\textbf{Root Tweet} & Video footage of WHITE people fliping over a cop car in SLC, Utah that later caught on fire. Our protest was peaceful. Black organizers wanted to remain peaceful. White people are responsible.[LINK]\\
\hline
\textbf{Reply Tweet} & @USER Like half of the people flipping the car are black, what the fuck are you talking about?\\
\hline
\hline
\textbf{Prompt} & You are analyzing a tweet on Twitter.: tweet: [ROOT\_TWEET] Without any explanation, provide three adjectives representing different sentimental qualities of the tweet. For every adjective you intend to provide, use their simplest form in English: \textbackslash n adjective 1: \textbackslash n adjective 2: \textbackslash n adjective 3:\\
\hline
\textbf{Response\newline Mistral-7B} &  \textbf{agitated:} The tweet expresses a sense of agitation towards the actions of the white people in the situation. \textbf{peaceful:} The tweet emphasizes that the protest was peaceful, suggesting a calming or soothing quality. \textbf{responsible:} The tweet assigns responsibility to the white people, implying a sense of seriousness or gravity.\\
\hline
\hline
\textbf{Prompt} & You are analyzing an interaction between two people on Twitter. User A has posted a tweet and User B has replied to it: User A: [ROOT\_TWEET] User B: [REPLY\_TWEET]. Without any explanation, describe the stance of user B's reply toward user A's tweet with three adjectives. Make sure that the adjectives you provide are with regards to the stance that is directed toward user A and/or his/her opinion, not toward an external concept or person. For every adjective you intend to provide, use their simplest form in English: \textbackslash n adjective 1: \textbackslash n adjective 2: \textbackslash n adjective 3:\\
\hline
\textbf{Response\newline Mistral-7B} &  \textbf{1. Skeptical:} User B expresses doubt about the accuracy of User A's statement that only white people were flipping the cop car. \textbf{2. Confrontational:} User B challenges User A's assertion directly and forcefully. \textbf{3. Curious:} User B seems interested in obtaining more information and clarification about the event described in User A's tweet.\\ \hline

\end{tabular}
\caption{Sample prompt and response for LLM-aided annotation of interactions}
\label{tab:prompt-sample}
\end{table*}

\begin{table}[ht]
\centering

\begin{tabular}{|c|p{6cm}|p{6cm}|c|}
\hline
\textbf{Stance} & \textbf{Root Tweet}                                                                                                                                                                                                          & \textbf{Reply Tweet}                                                                             & \textbf{Interaction} \\ \hline
patriotic       & I’m Proud To Be An American!                                                                                                                                                                                                 & @USER Love the song. Happy tears of pride.                                                       & $R \rightarrow R$    \\ \hline
Prayerful       & The Lord will make you great..                                                                                                                                                                                               & @USER The Lord will make me great. Amen.                                                         & $R \rightarrow R$    \\ \hline
Conspiratorial  & Giuliani Rips Fauci, Says US Paid for 'Damn Virus That's Killing Us' {[}LINK{]}                                                                                                                                              & @USER The Deep State at work with their cronies \#CrooksandCommunists                            & $R \rightarrow R$    \\ \hline
certain         & You Vote: Do you think Antifa is a domestic terrorist organization? | Just The News {[}LINK{]}                                                                                                                               & @USER Yep without a doubt                                                                        & $R \rightarrow R$    \\ \hline
unsupportive    & Your support means everything. I will never stop fighting for our movement. {[}LINK{]}                                                                                                                                       & @USER You don't have my support.                                                                 & $R \rightarrow D$    \\ \hline
imaginative     & If life were a 90s Fox drama, Harry and Meghan would abandon the royalty to start a fresh ad agency in LA.                                                                                                                   & @USER Set it at Christmas and you have a Hallmark movie in the making                            & $R \rightarrow D$    \\ \hline
proactive       & The Pope says tax evaders have stolen from the Government and weakened Italy's health scheme and are murderers {[}LINK{]}                                                                                                    & @USER We need to stop cash in hand as well, it's rife, especially in London \#LockdownNow        & $R \rightarrow D$    \\ \hline
\hl{congratulatory}       & Oh my gosh you guys... I have a son!!! This is crazy!! {[}IMAGE{]} & @USER Congratulations! Enjoy!        & $R \rightarrow D$    \\ \hline
\hl{loving}       & My baby is back home!! {[}IMAGE{]}  & @USER That's why you put that dress on..aaahhh  {[}IMAGE{]} & $R \rightarrow D$    \\ \hline
\hl{apologetic}       & This version of Hamburglar has definitely burned a church in Norway {[}LINK{]}  & @USER ``Sorry for partying''-the hamburglar probably & $R \rightarrow D$    \\ \hline
reciprocal      & Shots from the PEACEFUL \#BlackLivesMatter protest in Austin today. City made me proud! {[}LINK{]}                                                                                                                           & @USER I was there too. Incredibly hopeful to see so many people show up for change. \#BLMprotest & $D \rightarrow D$    \\ \hline
similar         & Admit it. You wish it was Sunday already. \#KillingEve                                                                                                                                                                       & @USER Oh my god yes .. I'm obsessed too                                                          & $D \rightarrow D$    \\ \hline
relatable       & I'm at work. I'm always at work. I'm losing my mind.                                                                                                                                                                         & @USER Oh good, I was starting to think I was the only one                                        & $D \rightarrow D$    \\ \hline
fascinated      & \#WhenTheLockdownEnds I'm going to party like a Maya Ruler in a weird lobster costume! {[}LINK{]}                                                                                                                            & @USER Wow! Is that from Bonampak? What a party!                                                  & $D \rightarrow D$    \\ \hline
empathetic      & RIP Grandpa.... We lost him to COVID-19 last night                                                                                                                                                                           & @USER I’m so sorry                                                                               & $D \rightarrow D$    \\ \hline
accusatory      & Why bother to tweet about this, of all things? {[}LINK{]}                                                                                                                                                                    & @USER He’s happy to have people dying to improve his ratings.                                    & $D \rightarrow R$    \\ \hline
hostile         & This is critical evidence confirming what we already knew to be true — China lied. China continues to lie. China must be held responsible {[}LINK{]}                                                                         & @USER shut up, traitor                                                                           & $D \rightarrow R$    \\ \hline
dismissive      & Trump, reading, says COVID-19, then adds, "You know what that is? Right. Become a very famous term. C-O-V-I-D. COVID."                                                                                                       & @USER He's a child.                                                                              & $D \rightarrow R$    \\ \hline
skeptical       & New: Biden says he will choose his running mate next week - CNNPolitics {[}LINK{]}                                                                                                                                           & @USER He said this immediately after becoming the nominee.                                       & $D \rightarrow R$    \\ \hline
critical        & President Trump says a new, faster coronavirus test is going to be used soon. He says: I hope the new test works out. Hopefully it will check out or test out. It will be a very simple test. It won't be unpleasant at all. & @USER Trump doing the right thing after exhausting all other options.                            & $D \rightarrow R$    \\ \hline
\end{tabular}
\caption{Sample tweets for AI-generated stances for replies.}
\label{tab:samples-stance}
\end{table}

\begin{table*}[ht]
\centering
\begin{tabular}{|l|p{12.5cm}|l|}
\hline
\textbf{Sentiment} & \textbf{Root Tweet}  & \textbf{Party} \\ \hline
significant        & Meanwhile, in other news, this important summit meeting to develop strategies to fight the pandemic.  & Democrat       \\ \hline
shameful           & There is no greater embarrassment in the House of Representatives than Jim Jordan.                                                                                                                                              & Democrat       \\ \hline
moral              & Anyone who thinks they need to go forward with an execution in this moment shouldn’t ever have the authority to carry out executions.                                                                                           & Democrat       \\ \hline
alert              & We are making progress. Some good recovery numbers are coming in. GHS should be announcing soon. Doesn’t mean we can let down our guard and live without care. Let’s keep observing the preventive etiquette.                   & Democrat       \\ \hline
perspective        & To people complaining about the wrong statues getting removed: if the right statues had been removed earlier or NEVER placed, you could have avoided all this.                                          & Democrat       \\ \hline
responsible        & If we’re going to rise to this moment with the attention and action that it fully deserves, we better start listening. We better own up to our own responsibility that led to this moment. And we better start inviting change. & Democrat       \\ \hline
proactive          & NEW: New York State will require all hospitals to have on hand a 90-day supply of PPE at quantities sufficient to meet the rate of use during the worst of this crisis.                                                         & Democrat       \\ \hline
balanced           & The level of anger directed at the media from these protestors was alarming. As always, I will tell a fair and unbiased story today.                                                                    & Democrat       \\ \hline
overwhelmed        & I cannot work. The Christmas break has broken me                                                                                                                                                                                & Democrat       \\ \hline
humorous           & My wife is a teacher. There's herding cats, and then there's "webinar with 20 1st graders"                                                                                                                                      & Democrat       \\ \hline
intimate           & My mom just passed away in her sleep. Does not appear covid related. I'm numb.                                                                                                                                                  & Democrat       \\ \hline
gentle             & Good morning sweet girl\textbackslash{}nHave a great day.                                                                                                                                                                       & Democrat       \\ \hline
unempathetic       & She can stand in the unemployment line..                                                                                                                                                                                        & Democrat       \\ \hline
unconventional     & There are no rules... for breakfast I just had garlic cheese bread along with my berry protein shake \textbackslash{}nWhatever                                                                                                  & Democrat       \\ \hline
tired              & I need to be in bed. \#QuarantineLife                                                                                                                                                                                           & Democrat       \\ \hline
surreal            & Even the bedbugs are wearing masks.                                                                                                                                                                                             & Democrat       \\ \hline
isolated           & Am I the only person who’s never watched Friends?                                                                                                                                                                               & Democrat       \\ \hline
misleading         & The President is rambling, reading a script full of badly formed theories. And lies.                                                                                                                                            & Republican     \\ \hline
interesting        & This is the most bizzare part of a fascinating thread about how Labour responded to the 1992 defeat.{[}link{]}                                                                                                                  & Republican     \\ \hline
arrogant           & Trump: ``We\'re doing a job the likes of which nobody's ever done.''                                                                                                                                                               & Republican     \\ \hline
radical            & AI is more dangerous than nuclear weapons.                                                                                                                                                                                      & Republican     \\ \hline
diplomatic         & US Secretary of State Mike Pompeo in his opening remarks in \#Doha says that the Taliban have committed to cut their ties with terrorist groups. \#Afghanistan \#PeaceDeal {[}link{]}                                           & Republican     \\ \hline
formal             & President Trump announces intent to appoint Dr. Sebastian Gorka, of Virginia, to be a Member of the National Security Education Board for a term of four years.                                                                 & Republican     \\ \hline
self-centered      & "We understand there are still many sick and dying...But what about the rest of us?" {[}link{]}                                                                                                                                 & Republican     \\ \hline
alarmist           & Our national media is ROOTING for hydroxycloroquine to not work as a treatment for \#Covid\_19. Think about that.                                                                                                               & Republican     \\ \hline
amused             & Just got a PR pitch that includes the word ``covidpreneur''                                                                                                                                                                       & Republican     \\ \hline
accusatory         & The Chinese Communist Party is responsible for every single death, every job lost, every retirement nest egg lost, from this coronavirus. And Xi Jinping must be made to pay the price.                                         & Republican     \\ \hline
negative           & A rebound in 2021 will not happen if we keep doing this... {[}link{]}                                                                                                                                   & Republican     \\ \hline
disgusted          & What kind of man uses a funeral to push a political agenda? {[}link{]}                                                                                                                                                          & Republican     \\ \hline
outraged           & Wake the f*** up America! {[}link{]}                                                                                                                                                                                            & Republican     \\ \hline
sarcastic          & A face mask is part of the 'scamdemic,' they say. But they'll be happy to sell you one {[}link{]}                                                                                                                               & Republican     \\ \hline
prayerful          & Please Lord let it be so. {[}link{]}                                                                                                                                                                                       & Republican     \\ \hline
prejudiced         & Black people love acting like kings, queens and African warriors, but never like normal Americans. {[}link{]}                                                                                                                   & Republican     \\ \hline
conspiratorial     & The World your mind has been MK Ultra Programmed into believing you live in, bears no resemblance to the World you are actually living in.                                                                                      & Republican     \\ \hline
\end{tabular}
\caption{Sample tweets for AI-generated root sentiments.}
\label{tab:samples-root}  
\end{table*}

\end{document}